\documentclass[letterpaper,12pt]{JHEP3}
\usepackage{amsmath}
\usepackage{amssymb}
\usepackage{epsfig}
\newcommand{\labell}[1]{\label{#1}}
\def\({\left(} \def\){\right)}
\def\[{\left[} \def\]{\right]}
\def\al{\alpha} \def\bt{\beta}
\def\del{{\partial}}
\newcommand{\non}{\nonumber \\}

\newcommand{\be}{\begin{equation}}
\newcommand{\ee}{\end{equation}}
\newcommand{\bea}{\begin{eqnarray}}
\newcommand{\eea}{\end{eqnarray}}
\newcommand{\ba}{\begin{eqnarray}}
\newcommand{\ea}{\end{eqnarray}}

\newcommand{\beq}{\begin{equation}}
\newcommand{\eeq}{\end{equation}}
\newcommand{\beqa}{\begin{eqnarray}}
\newcommand{\eeqa}{\end{eqnarray}}
\newcommand{\beqar}{\begin{eqnarray*}}
\newcommand{\eeqar}{\end{eqnarray*}}

\newcommand{\reef}[1]{(\ref{#1})}

\newcommand{\eg}{{\it e.g.,}\ }
\newcommand{\ie}{{\it i.e.,}\ }

\newcommand{\mt}[1]{\textrm{\tiny #1}}

\newcommand{\ga}{\gamma}

\newcommand{\A}{\mathcal{A}}

\newcommand{\cO}{\mathcal{O}}
\newcommand{\R}{\mathcal{R}}

\newcommand{\la}{\lambda}
\newcommand{\lp}{\ell_{\mt P}}

\newcommand{\hi}{{\hat \imath}}
\newcommand{\hj}{{\hat \jmath}}

\newcommand{\cA}{{\cal A}}
\newcommand{\cV}{{\cal V}}

\def\hz{\mathrel{\mathop h^{\scriptscriptstyle{(0)}}}{}\!\!}
\def\Xz{\mathrel{\mathop X^{\scriptscriptstyle{(0)}}}{}\!\!}
\def\Xo{\mathrel{\mathop X^{\scriptscriptstyle{(1)}}}{}\!\!}
\def\Xs{\mathrel{\mathop X^{\scriptscriptstyle{(2)}}}{}\!\!}

\def\gz{\mathrel{\mathop g^{\scriptscriptstyle{(0)}}}{}\!\!\!}
\def\go{\mathrel{\mathop g^{\scriptscriptstyle{(1)}}}{}\!\!\!}
\def\gs{\mathrel{\mathop g^{\scriptscriptstyle{(2)}}}{}\!\!\!}
\def\gn{\mathrel{\mathop g^{\scriptscriptstyle{(n)}}}{}\!\!\!}
\def\hz{\mathrel{\mathop h^{\scriptscriptstyle{(0)}}}{}\!}
\def\ho{\mathrel{\mathop h^{\scriptscriptstyle{(1)}}}{}\!}

\newcommand{\Xf}[1]{\mathrel{\mathop X^{\scriptscriptstyle{(#1)}}}{}\!\!}
\newcommand{\Yf}[1]{\mathrel{\mathop Y^{\scriptscriptstyle{(#1)}}}{}\!\!}
\newcommand{\gf}[1]{\mathrel{\mathop g^{\scriptscriptstyle{(#1)}}}{}\!\!\!}
\newcommand{\hf}[1]{\mathrel{\mathop h^{\scriptscriptstyle{(#1)}}}{}\!\!\!}
\newcommand{\fff}[1]{\mathrel{\mathop f^{\scriptscriptstyle{(#1)}}}{}\!\!\!\!}
\newcommand{\XX}[1]{\mathrel{\mathop X^{\scriptscriptstyle{(#1)}}}{}\!\!\!\!}

\newcommand{\m}{v}
\newcommand{\Sig}{{\partial V}}

\preprint{arXiv:1105.nnnn [hep-th]}

\title{Some Calculable Contributions to Holographic Entanglement Entropy}

\author{Ling-Yan Hung, Robert C. Myers
and Michael Smolkin\\
{\it Perimeter Institute for Theoretical Physics,
\\ 31 Caroline Street North, Waterloo,
Ontario N2L 2Y5, Canada}}

\abstract{Using the AdS/CFT correspondence, we examine entanglement entropy
for a boundary theory deformed by a relevant operator and establish two
results. The first is that if there is a contribution which is
logarithmic in the UV cut-off, then the coefficient of this term is
independent of the state of the boundary theory. In fact, the same is
true of all of the coefficients of contributions which diverge as some
power of the UV cut-off. Secondly, we show that the relevant
deformation introduces new logarithmic contributions to the
entanglement entropy. The form of some of these new contributions is similar to that
found recently in an investigation of entanglement entropy in a free massive
scalar field theory \cite{frank}.}

\begin{document}

\section{Introduction}

Entanglement entropy was first considered in the context of the AdS/CFT
correspondence by Ryu and Takayanagi \cite{rt1,rt2}. They provided a
simple conjecture for calculating holographic entanglement entropy. In
the $d$-dimensional boundary field theory, the entanglement entropy
between a spatial region $V$ and its complement $\bar V$ is given by
the following expression in the ($d$+1)-dimensional bulk spacetime:
 \be
S(V) = \frac{2\pi}{\lp^{d-1}}\ \mathrel{\mathop {\rm
ext}_{\scriptscriptstyle{\m\sim V}} {}\!\!} \left[A(\m)\right]
 \labell{define}
 \ee
where $\m\sim V$ indicates that $\m$ is a bulk surface that is
homologous to the boundary region $V$ \cite{head,furry}. In particular,
the boundary of $\m$ matches the `entangling surface' $\partial V$ in
the boundary geometry. The symbol `ext' indicates that one should
extremize the area over all such surfaces $\m$.\footnote{We are using
`area' to denote the ($d$--1)-dimensional volume of $\m$. If
eq.~\reef{define} is calculated in a Minkowski signature background,
the extremal area is only a saddle point. However, if one first Wick
rotates to Euclidean signature, the extremal surface will yield the
minimal area.} Implicitly, eq.~\reef{define} assumes that the bulk
physics is described by (classical) Einstein gravity and we have
adopted the convention: $\lp^{d-1}=8\pi G_\mt{N}$ Hence we may observe
the similarity between this expression \reef{define} and that for black
hole entropy. While this proposal passes a variety of consistency
tests, \eg see \cite{head,rt2,EEGB}, there is no general derivation of
this holographic formula \reef{define}. However, a derivation has
recently been provided for the special case of a spherical entangling
surface in \cite{circle4}.

One aspect of entanglement entropy (EE), which eq.~\reef{define}
reproduces, is that this quantity diverges and so it is only
well-defined with the introduction of a short-distance cut-off $\delta$
in the boundary field theory. Generically, the leading term obeys an
`area law,' being proportional to $\A_{d-2}/\delta^{d-2}$ where
$\A_{d-2}$ denotes the area of the entangling surface in the boundary
theory. While the coefficient of this divergent contribution is
sensitive to the details of the UV regulator, universal data
characterizing the underlying field theory can be found in subleading
contributions. In particular, in a conformal field theory (CFT) with
even $d$, one typically finds a logarthmic term $\log(R/\delta)$ where
$R$ is some macroscopic scale characterizing the size of the region
$V$. The coefficient of this logarithmic contribution is a certain
linear combination of the central charges appearing in the trace
anomaly of the CFT. The precise details of the linear combination will
depend on the geometry of the background spacetime and of the
entangling surface \cite{EEGB,rt2,solo,cthem2}.

A similar class of universal contributions were recently identified in
a calculation with a free massive scalar field
\cite{frank}.\footnote{See \cite{reallyold} for related results in
$d=4$.} In particular, considering the scalar field theory (in a
`waveguide' geometry) with an even number of spacetime dimensions, a
logarithmic contribution to the EE appears with the form
 \be
S_\mt{univ}=-\gamma_d\,\cA_{d-2}\,\mu^{d-2}\,\log(\mu\delta)\,,
 \labell{franks}
 \ee
where $\mu$ is the mass of the scalar and $\gamma_d$ is a numerical
factor depending on the spacetime dimension.\footnote{To be precise,
$\gamma_d=(-)^{\frac{d-2}{2}} [6(4\pi)^{\frac{d-2}{2}}
\Gamma(d/2)]^{-1}$ \cite{frank}.} While this calculation is simplified
by having a free field theory, one would still characterize the mass
term as being a relevant operator. That is, the mass dominates the
physics of the scalar theory at low energies but it leaves the leading
UV properties unchanged, \eg the area law contribution to the EE would
not be affected.

With this perspective, one is led to examine a natural strong coupling
analog of this calculation using holography.  In particular, with the
standard AdS/CFT dictionary \cite{revue}, we can introduce a relevant
deformation of the boundary field theory by turning on a (tachyonic)
scalar field in the bulk. Asymptotically the bulk geometry still
approaches an AdS spacetime reflecting the fact that the boundary
theory still behaves as a CFT in the far UV. However, the details of
the bulk geometry are changed by the back-reaction of the scalar field
and so this naturally introduces the possibility that applying
eq.~\reef{define} in this geometry will yield new universal
contributions of the form given above. In fact, our holographic
calculations reveal a general class of logarithmic contributions which
can appear with a relevant deformation. Schematically, they take the
form of an integral over the entangling surface $\Sig$:
 \be
S_\mt{univ} = \sum_{i,n} \gamma_i(d,n) \int_{\Sig}\!d^{d-2}\!\sigma\,
\sqrt{H}\ [R,K]^n_i\ \mu^{d-2-2n}\,\log \mu\delta\,, \labell{EEdisk2x}
 \ee
where $n<(d-2)/2$ , $\mu$ is the mass scale appearing in the coupling of the new
relevant operator, $H_{ab}$ is the induced metric on $\Sig$ and
$[R,K]^n_i$ denotes various combinations of the curvatures with a
combined dimension $2n$. Both the curvature of the background geometry
or the extrinsic curvature of the entangling surface may enter these
expressions. The coefficients $\gamma_i$ depend on the details of the
underlying field theory and provide universal information
characterizing this theory. For an even-dimensional CFT, only the
contributions with $n=(d-2)/2$ appear (and in the logarithm $\mu$ would
be replaced by a scale in the geometry) and the coefficients $\gamma_i$
are proportional to the central charges of the CFT \cite{rt2,solo}. Of
course, for $n=0$, there is a single contribution which matches that
appearing in eq.~\reef{franks}.

One feature which is typically implicit in calculations of entanglement
entropy is that the QFT is in its vacuum state. However, it is further
assumed that the coefficients appearing in these logarithmic
contributions are `universal,' including that they are
independent of the state of the underlying QFT. For example,
calculating the entanglement entropy in a thermal bath would yield
precisely the same result as in the vacuum state. While this feature is
relatively `obvious' and can be confirmed with explicit calculations, a
rigorous proof is lacking. The basic idea is that the properties of the
state will not modify the UV properties of the theory. One of our
results in the following will be to demonstrate that the logarithmic
contributions are in fact independent of the state in a
holographic setting. Further our analysis makes clear that the
coefficient of these contributions is a local functional of the
geometry of the background in which the boundary theory resides and of
the entangling surface.

An overview of the paper is as follows: We begin in section \ref{notes}
by demonstrating  that the coefficient of any logarithmic contribution
is independent of the state of the boundary theory. Our first
discussion in this section considers the boundary theory being a pure
CFT but in section \ref{rest}, we show that this result extends to the
case where the boundary theory is deformed by a relevant operator. In
fact our conclusion is that in general any UV divergent terms involve
local functionals of the geometry and couplings of the boundary theory.
An additional feature which our analysis reveals is that new universal
contributions to the entanglement entropy can arise from the presence
of the relevant deformation. Hence in the subsequent sections, we
present some explicit examples where such logarithmic contributions are
calculated. We begin in section \ref{exercise} by considering flat and
spherical entangling surfaces with the boundary theory in a flat
background. This exercise demonstrates that, as well as terms of the
form \reef{franks}, there are also universal contributions where the
mass scale of the relevant deformation combines with the curvature of
the entangling surface, as shown in eq.~\reef{EEdisk2x}. In section
\ref{curve}, we investigate the latter contributions further by
considering examples where the background in which the boundary theory
resides is curved, \eg $R\times S^{d-1}$ and $R\times H^{d-1}$. In
section \ref{pbhmatter}, we extend the approach of \cite{adam} to
properly identify the precise structure of these contributions in the
leading case. We conclude with a discussion of our brief results, in
section \ref{discuss}. Finally, in appendix \ref{oddu}, we present a
holographic calculation which explicitly shows that the constant terms,
which are often interpreted as universal, in odd dimensions do indeed
depend on the state of the boundary theory.

\section{Universality with a Boundary CFT} \label{notes}

In this section, we establish that the logarithmic contribution to the
entanglement entropy (EE) is independent of the state of the boundary
field theory in the AdS/CFT correspondence. To begin, we consider the
case where the boundary theory is purely a conformal field theory. We
must then also choose the boundary dimension to be
even,\footnote{However, our discussion here will consider both odd and
even $d$ because both are germane when a relevant deformation is
introduced in the next subsection. We comment further on the case of
odd $d$ in appendix \ref{oddu}.} since for a CFT, it is only in this
case that a logarithmic contribution arises in the EE. It has been
shown that the coefficient of this term is related to the central
charges appearing in the trace anomaly \cite{finn,rt2,solo} --- see
also the discussion in \cite{cthem2}. Implicitly, the calculations
establishing this connection are made in the vacuum of the
corresponding CFT and so here we are establishing that, at least in a
holographic framework, the result is independent of the state of the
CFT. Our present observation comes as a simple extension of the
holographic calculations in \cite{EEGB}. There our holographic
calculations were able to reproduce the precise expression for the
logarithmic term in the EE for a general entangling surface in a
four-dimensional CFT, matching that which was originally determined in
\cite{solo}.

Let us begin by denoting the spacetime dimension of the boundary theory
as $d$ and hence the dual gravity theory has $d+1$
dimensions.\footnote{Our notation is essentially the same as that
established in \cite{EEGB}. Explicitly then, our index conventions are
as follows: Directions in the full (AdS) geometry are labeled with
letters from the second half of the Greek alphabet, \ie
$\mu,\nu,\rho,\cdots$. Letters from the `second' half of the Latin
alphabet, \ie $i,j,k,\cdots$, correspond to directions in the
background geometry of the boundary CFT. Frame indices are denoted by a
hat, \ie $\hi,\hj$. Meanwhile, directions along the entangling surface
in the boundary are denoted with letters from the beginning of the
Latin alphabet, \ie $a,b,c,\cdots$, and directions along the
corresponding bulk surface are denoted with letters from the beginning
of the Greek alphabet, \ie $\al,\bt,\ga,\cdots$. \label{footy}} In the
AdS/CFT correspondence, the bulk geometry asymptotically approaches
anti-de Sitter space for any generic state of the boundary theory.
This asymptotic geometry can then be described with the
Fefferman-Graham expansion, as follows \cite{feffer} --- see also
\cite{construct}:
 \be
ds^2 = \frac{L^2}{4}\frac{d\rho^2}{\rho^2} + \frac{1}{\rho}\,g_{i
j}(x,\rho)\,dx^i dx^j\,,\labell{expandfg}
 \ee
where $L$ is the AdS curvature scale. The asymptotic boundary is
approached with $\rho\to0$ and $g_{i j}(x,\rho)$ admits a series
expansion in the (dimensionless) radial coordinate $\rho$
 \bea
g_{i j}(x,\rho)&=& \gz_{i j}(x^i) + \rho \go_{i j}(x^i) + \rho^2 \gs_{i
j}(x^i) \labell{expand}\\
&& \qquad + \cdots + \rho^{d/2} \gf{d/2}_{i j}(x^i) +
\rho^{d/2}\log\rho\,\fff{d/2}_{ij}(x^i)+O(\rho^{\frac{d+1}2})\,.
 \nonumber
 \eea
The leading term $\overset{\scriptscriptstyle{(0)}}{g}\!_{ij}$
corresponds to the metric on which the boundary CFT resides. As shown
in eq.~\reef{expand}, the first few terms fall into a Taylor series
expansion but this simple form breaks down at order $\rho^{d/2}$. In
particular, for even $d$, a logarithmic term appears at this order
while for odd $d$, non-integer powers of $\rho$ begin to make an
appearance --- no logarithmic term appears for odd $d$. With this
choice of coordinate system, the expectation value of the boundary
stress-energy tensor becomes \cite{construct,nice}
 \be
\langle\, T_{ij}\,\rangle = \frac{d}{2\,\lp^{d-1}L}\gf{d/2}_{ij} +
\widetilde{X}_{ij}[\,\gn\,\,] \,. \labell{stress}
 \ee
where $\widetilde{X}_{ij}$ denotes the contribution coming from the
conformal anomaly. Hence this term vanishes for odd $d$, while for even
$d$, it is determined by coefficients
$\overset{\scriptscriptstyle{(n)}}{g}\!_{ij}$ with $n < d/2$.

In solving the bulk Einstein equations, both
$\overset{\scriptscriptstyle{(0)}}{g}\!_{ij}$ and
$\overset{\scriptscriptstyle{(d/2)}}{g}\!\!_{ij}$ can be regarded as
the independent boundary data needed to determine the bulk spacetime.
As noted above, the first fixes the boundary metric while the second
determines the boundary stress tensor. That is,
$\overset{\scriptscriptstyle{(d/2)}}{g}\!\!_{ij}(x)$ is related to the
state of the boundary CFT. Further, we note that the coefficients
$\overset{\scriptscriptstyle{(n)}}{g}\!_{ij}(x)$ with $0<n<d/2$ are
completely fixed in terms of the boundary metric
$\overset{\scriptscriptstyle{(0)}}{g}\!_{ij}$. More precisely, by
expanding the gravitational equations of motion near the boundary, one
solves for each of these coefficients in terms of the lower order terms
in the expansion \reef{expand} --- see section \ref{rest} for more
details. For example (as long as $d>2$), one finds \cite{adam}:
 \be
\go_{i j} = -\frac{{L}^2}{d-2}\bigg(R_{i j}[\gz\,]-
\frac{\overset{\scriptscriptstyle{(0)}}{g}_{ij}}{2(d-1)}R[\gz\,]
\bigg)\,,
 \labell{metricexpand}
 \ee
where $R_{ij}$ and $R$ are the Ricci tensor and Ricci scalar
constructed with the boundary metric
$\overset{\scriptscriptstyle{(0)}}{g}\!_{ij}$. An alternative approach
was presented in \cite{adam}. There the authors showed that these
coefficients are almost completely fixed by conformal symmetries at the
boundary.\footnote{These calculations leave some small ambiguity that
must still be fixed by the equations of motion.} This method examines
the effect of Penrose-Brown-Henneaux (PBH) transformations, the
subgroup of bulk diffeomorphisms which generate Weyl transformations of
the boundary metric. Consistency of the PBH transformations on the
asymptotic expansion \reef{expand} essentially determines all of the
coefficients up to order $n<d/2$ --- see section \ref{pbhmatter} for
further details. This approach also makes clear that these coefficients
can be expressed as covariant tensors constructed from curvatures of
the boundary metric (as well as covariant derivatives of these), as
illustrated in eq.~\reef{metricexpand}. Further, one finds that the
resulting expression for
$\overset{\scriptscriptstyle{(n)}}{g}\!_{ij}(x)$ contains $2n$
derivatives.

As these coefficients in the asymptotic geometry \reef{expand} depend
only on the boundary metric
$\overset{\scriptscriptstyle{(0)}}{g}\!_{ij}$ and are completely
independent of the state of the boundary CFT, we refer to them as the
`fixed boundary data.' Hence, if the logarithmic contribution to the EE
is independent of the state, we must demonstrate that in our
holographic calculations of the EE (for even $d$), this contribution
depends only on this fixed boundary data and is independent of any
coefficients $\overset{\scriptscriptstyle{(n)}}{g}\!_{ij}$ with $n\ge
d/2$.

The holographic EE is determined by evaluating eq.~\reef{define} and
hence the structure of the result, \ie the coefficient of the
logarithmic contribution, depends on the geometry of the extremal
surface $\m$. Hence, as well as the bulk geometry \reef{expand}, we
must also consider the embedding of the corresponding
($d-1$)-dimensional surface in the $(d+1)$-dimensional bulk geometry.
This embedding may be described by $X^\mu=X^\mu(y^{a},\tau)$, where
$X^\mu=\lbrace x^i,\rho\rbrace$ are the bulk coordinates and
$\sigma^\al= \lbrace y^a,\tau\rbrace$ are the coordinates on surface
$m$ (with $a=1,..,d-2$) -- recall our conventions from footnote
\ref{footy}. The induced metric on the bulk surface is then given by
 \be
h_{\al\bt}=\del_\al X^\mu \del_\bt X^\nu\, g_{\mu\nu}[X]\,,
 \labell{induceh}
 \ee
where $g_{\mu\nu}$ denotes the full $(d+1)$-dimensional bulk metric.

The calculations below are simplified somewhat if we fix
reparameterizations of the coordinates on $\m$ with the following gauge
choices (as in \cite{adam})
 \be
 \tau=\rho\qquad{\rm and}\qquad h_{a\tau}=0
 \, .
 \labell{eqn:gauge-brane}
 \ee
Now following \cite{grwit}, one finds that the remaining embedding
functions $X^i(y^{a},\tau)$ are then described by the following series
expansion\footnote{The gauge condition \reef{eqn:gauge-brane} is not
the same as in \cite{grwit}, however, the general structure of the
asymptotic expansion \reef{expandx} remains unaltered in both cases.}
for small $\rho$:
  \bea
X^i(y^a,\rho)&=& \Xz^i(y^a) + \tau \Xo^i(y^a) + \tau^2 \Xs^i
(y^a) \labell{expandx}\\
&& \qquad + \cdots + \tau^{d/2} \Xf{d/2}^{i}(y^a) +
\tau^{d/2}\log\tau\,\Yf{d/2}^{i}(y^a)+O(\tau^{\frac{d+1}2})\,.
 \nonumber
 \eea
Essentially the form of this expansion matches that for the bulk metric
in eq.~\reef{expand}. The first term
$\overset{\scriptscriptstyle{(0)}}{X^i}(y^a)$ describes the position of
$\del \m$ in the boundary of the asymptotically AdS space. That is,
this matches the position of the entangling surface $\Sig$ in the
boundary metric $\overset{\scriptscriptstyle{(0)}} g_{ij}(x)$. Here as
in eq.~\reef{expand}, the simple Taylor series expansion appearing for
the first few terms breaks down at order $\rho^{d/2}$. Again, for even
$d$, a logarithmic term appears at this order while for odd $d$,
non-integer powers of $\rho$ begin to appear --- no logarithmic term
appears for odd $d$. This expansion \reef{expandx} is constructed in
detail by solving the local equations of motion for $X^\mu(y^{a},\tau)$
which extremize the area of the bulk surface $\m$ \cite{grwit} --- see
section \ref{rest} for more details. In solving these equations, the
full surface is determined by independently fixing both
$\overset{\scriptscriptstyle{(0)}}{X^i}(y^a)$ and
$\overset{\scriptscriptstyle{(d/2)}}{X^i}(y^a)$. In particular, the
latter data would be chosen to ensure that the surface $\m$ closes off
smoothly in the interior of the asymptotically AdS space. As the
equations are solved iteratively order by order in $\tau$, the
coefficients $\overset{\scriptscriptstyle{(n)}}{X^i}(y^a)$ with $n<d/2$
are completely determined as local functionals of
$\overset{\scriptscriptstyle{(0)}}{X^i}(y^a)$ and
$\overset{\scriptscriptstyle{(0)}}{g}_{ij}$. In particular then, these
terms are independent of the state of the boundary CFT. Hence in
discussing the extremal surface $m$, we extend the meaning of the
`fixed boundary data' to include both
$\overset{\scriptscriptstyle{(n)}}{g}_{ij}$ and
$\overset{\scriptscriptstyle{(n)}}{X^i}(y^a)$ with $n<d/2$.

As a further note, we add that these leading contributions for the
embedding functions can again be determined with the application of PBH
transformations \cite{adam} --- see section \ref{pbhmatter} for more
details. We also comment that the analysis in refs.~\cite{grwit,adam}
was more general in considering extremal bulk submanifolds ending on a
boundary surface with an arbitrary dimension $k$. In general, they
found that the second set of independent coefficients entered the
expansion of the embedding functions at order $\tau^{(k+2)/2}$. Hence
it is only for $k=d-2$, the case of present interest, that the form of
the expansion \reef{expandx} matches the metric expansion
\reef{expand}.

Given the expansions of the bulk metric \reef{expand} and the embedding
functions \reef{expandx}, the induced metric \reef{induceh}, compatible
with the gauge choice (\ref{eqn:gauge-brane}), can also be expanded in
the vicinity of the AdS boundary
 \be
h_{\tau\tau}={L^2\over4\tau^2}\Big(1+\ho_{\tau\tau}
 \,\tau+\cdots\Big)\,,\qquad
 h_{ab}={1\over \tau}\(\hz_{ab}+\ho_{ab}\,\tau+\cdots\)\,.
\labell{inducemet}
 \ee
Note that $\overset{\scriptscriptstyle{(0)}}{h}_{ab}=H_{ab}$, \ie it
is precisely the induced metric on the entangling surface $\Sig$ as
evaluated in the boundary CFT. A crucial feature emerging from this
perturbative construction is that the coefficients
$\overset{\scriptscriptstyle{(n)}}{h}_{\alpha\beta}$ depend only on the
fixed boundary data for $n<d/2$, \ie they are completely determined by
$\overset{\scriptscriptstyle{(0)}}{X^i}(y^a)$ and
$\overset{\scriptscriptstyle{(0)}}{g}_{ij}$. Now in calculating the
holographic EE \reef{define}, we must evaluate the area
 \be
A(\m)=\int_\m d^{d-1}\sigma \sqrt{h} =\int_\m d^{d-2}y\, d\tau
\frac{L}{2\tau^{d/2}}\sqrt{\det\hz}\left[1 +\left(\ho_{\tau\tau} +
\hz^{ab}\,\ho_{ab} \right)\frac{\tau}2+\cdots\right]\,.
 \labell{arealx}
 \ee
The integral over the radial direction $\tau$ extends down to an
asymptotic regulator surface at $\tau_{min}=\rho_{min}=\delta^2/L^2$
where $\delta$ is a short distance cut-off in the boundary theory. We
are interested in the appearance of a logarithmic contribution of the
form $\log\delta$, hence we must carry out the expansion in the
bracketed expression to order $\tau^{\frac{d-2}2}$, which produces the
term in the integral with an overall power $1/\tau$. The explicit term
appearing at this order for general $d$ would be quite complicated.
However, for our purposes, it suffices to know that this term will
involve coefficients $\overset{\scriptscriptstyle{(n)}}{h}_{\al\bt}$
with $n\le(d-2)/2$. Hence this logarithmic contribution to the
holographic EE is completely determined by the fixed boundary data.
That is, we do not require the state dependent coefficient
$\overset{\scriptscriptstyle{(d/2)}}{g}_{ij}$, nor details of the shape
of the extremal surface $\m$ beyond what is determined by the boundary
geometry $\overset{\scriptscriptstyle{(0)}}{g}_{ij}$ and
$\overset{\scriptscriptstyle{(0)}}{X^i}(y^a)$. We also observe that, as
expected, there is no term at the appropriate order in
eq.~\reef{arealx} to produce a logarithmic contribution in the case of
odd $d$. Further, our analysis above shows that all of the divergent
contributions to the holographic EE will only depend on this fixed
boundary data, \ie these contributions are completely determined as
local functionals of $\overset{\scriptscriptstyle{(0)}}{g}_{ij}$ and
$\overset{\scriptscriptstyle{(0)}}{X^i}(y^a)$.

\subsection{Universality with a Relevant Deformation} \label{rest}

Motivated by the recent results in \cite{frank}, we wish to consider
how the holographic EE is modified by the introduction of a mass term
in the boundary theory. For example, a scalar mass term would deform
the boundary theory by an operator of dimension $\Delta=d-2$ while a
fermion mass term would introduce a deformation of dimension
$\Delta=d-1$. Our analysis here will be more general and consider
modifications of holographic EE when the boundary theory is deformed by
a general relevant (scalar) operator with $\Delta<d$. The dual of such
a scalar operator will be a scalar field in the bulk. Hence our
starting point is the following bulk action where we have Einstein
gravity coupled to a scalar field
 \beq
I = \frac{1}{2\lp^{d-1}} \int \mathrm{d}^{d+1}x \, \sqrt{-G}\, \left[ R
-\frac12(\partial\Phi)^2-V(\Phi)  \right]\,,
 \labell{action}
 \eeq
where
 \beq
V(\Phi)=-\frac{d(d-1)}{L^2}+\frac12 m^2\Phi^2+
\frac{\kappa}{6L^2}\Phi^3 + O(\Phi^4)\,.
 \labell{pot}
 \eeq

Now a relevant operator primarily affects the IR properties of the
theory but has an insignificant effect in the UV regime. In the present
holographic framework then, the dual bulk geometry still approaches AdS
space in the presence of the relevant operator. Hence we consider the
background geometry which asymptotically approaches AdS$_{d+1}$ in
Graham-Fefferman coordinates \cite{feffer}, as in eq.~\reef{expandfg},
 \be
ds^2=\frac{L^2}{4}\frac{d\rho^2}{\rho^2}+\frac{1}\rho \,g_{ij}(x,\rho)
\,dx^i dx^j\,.
 \labell{back0}
 \ee
Of course, with $\Phi=0$, the vacuum solution in the bulk will be
precisely AdS$_{d+1}$, as described in the previous section. In
general, the boundary theory's metric is still given by $g_{i
j}(x,\rho=0) = \overset{\scriptscriptstyle{(0)}}{g}\!_{ij}$, however,
as we will see below, the details of the small-$\rho$ expansion will
change with $\Phi\ne0$. If we turn on the scalar as a probe field in
this background, the scalar has two independent solutions
asymptotically \cite{revue}
 \be
 \Phi \simeq \rho^{\Delta_-/2}\phi^{(0)} +
 \rho^{\Delta_+/2}\phi^{(\Delta-\frac{d}2)}\,,
 \labell{field0}
 \ee
where
 \be
 \Delta_\pm=\frac{d}2\pm\sqrt{\frac{d^2}4+m^2L^2}\,.
 \labell{dimension}
 \ee
The standard approach is to interprete the conformal dimension of the
dual operator as $\Delta=\Delta_+$. Then, the leading coefficient of
the more slowly decaying solution, $\phi^{(0)}$, is interpreted as the
coupling for the dual operator in the boundary theory, while
$\phi^{(\Delta-\frac{d}2)}$ yields the expectation value of this
operator as \cite{igor9}\footnote{In general, additional contributions
may appear on the right-hand side involving $\phi^{(n)}$ with
$n<\Delta-\frac{d}2$. These terms are related to contact terms in
correlation functions of $\cO$ with itself and with the stress-energy
tensor \cite{construct}.}
 \be
\langle \cO(x)\rangle = (2\Delta-d)\,\phi^{(\Delta-\frac{d}2)}(x)\,.
 \labell{expect2}
 \ee
Since we wish to consider a relevant operator in the boundary theory,
eq.~\reef{dimension} requires that $m^2<0$ for the bulk scalar and
then, in fact, both of the modes in eq.~\reef{field0} decay
asymptotically (as $\rho\to0$). Note that this approach allows us to
study $\Delta\ge d/2$ and we must an `alternative quantization' to
study operators of lower dimension \cite{igor9}. We return to this
issue in the discussion in section \ref{discuss}.

While eq.~\reef{field0} was derived in the probe approximation, we will
see below that the powers appearing in the asymptotic scalar expansion
do not change when we consider the fully back-reacted solution. The
latter arises because we are considering a relevant operator with
$\Delta< d$.\footnote{In the case of a marginal operator with
$\Delta=d$, the boundary theory remains a CFT and so the results
regarding the holographic EE are unchanged from our previous
discussion.} Hence even in the full nonlinear analysis of the equations
of motion, $\Phi\to0$ as $\rho\to0$ and so the scalar remains a small
perturbation asymptotically. Hence the leading power in the
small-$\rho$ expansion of the scalar field is $\rho^{\alpha/2}$ with
$\alpha=\Delta_-$ and
 \be
0< \alpha\le\frac{d}2\,.
 \labell{range0}
 \ee
Here, the upper limit on $\alpha$ is the BF bound while the lower limit
is simple the requirement that $\Delta<d$.

Let us now turn to holography with the fully back-reacted solutions of
the Einstein-scalar theory \reef{action}. This construction is
discussed in some detail in \cite{construct} and we follow their
discussion below. The Einstein equations can be expressed as
 \beq
R_{\mu\nu}=\frac12\partial_\mu\Phi\partial_\nu \Phi+\frac1{d-1}
G_{\mu\nu} V(\Phi)\,,
 \labell{Einstein}
 \eeq
and the scalar wave equation is
 \beq
 \frac{1}{\sqrt{-G}}\partial_\mu\left(
\sqrt{-G} G^{\mu\nu}
\partial_\nu\Phi \right) - \frac{\delta V}{\delta\Phi}=0\,.
 \labell{scalar}
 \eeq
Now we write the scalar field as
 \be
\Phi(x,\rho)=\rho^{\alpha/2}\,\phi(x,\rho)
 \labell{scala}
 \ee
where we have extracted the leading asymptotic decay for this field.
Combining this form with the metric ansatz \reef{back0}, the Einstein
equations \reef{Einstein} yield \cite{construct}
 \bea
\rho\left(2g''_{ij}-2g'_{ik}g^{k\ell}g'_{\ell
j}\right.&+&\left.g^{k\ell}g'_{k\ell}\,g'_{ij}\right)+L^2R_{ij}[g]-(d-2)g'_{ij}
-g^{k\ell}g'_{k\ell}\, g_{ij}=
 \non
&=&\frac{\rho^{\alpha}}2\left(L^2\partial_i\phi\,\partial_j\phi
+\frac{g_{ij}}{(d-1)\rho}\left(m^2L^2\phi^2+\frac{\kappa}{3}\rho^{\alpha/2}
\phi^3+O(\rho^\alpha\phi^4)\right)\right)
 \non
\nabla_i\left(g^{k\ell}g'_{k\ell}\right)-\nabla^kg'_{ki}&=&\rho^{\alpha}L
\left(\phi'\,\partial_i\phi+\frac{\alpha}{2\rho}\phi\,\partial_i\phi\right)
 \labell{einsteinx}\\
g^{k\ell}g''_{k\ell}-\frac12 g^{ij}g'_{jk}g^{k\ell}g'_{\ell\,
i}&=&\rho^\alpha\left(\phi^{\prime\,2}+\frac{\alpha}{\rho}\phi\,\phi'+
\frac{\alpha^2}{4\rho^2}\phi^2
 \right.\non
&&\left.\qquad\qquad\ \ \
+\frac{1}{4(d-1)\rho^2}\left(m^2L^2\phi^2+\frac{\kappa}{3}
\rho^{\alpha/2} \phi^3+O(\rho^\alpha\phi^4)\right)
 \right)\nonumber
 \eea
where the primes denote differentiation with respect to $\rho$ and
$\nabla_i$ is the covariant derivative constructed from the metric
$g_{ij}(x,\rho)$. Further $R_{ij}[g]$ in the first line above denotes
the Ricci tensor calculated for the $d$-dimensional metric
$g_{ij}(x,\rho)$, treating $\rho$ as an extra parameter  --- in
particular then, this is {\it not} just the boundary Ricci tensor
calculated with $\overset{\scriptscriptstyle{(0)}}{g}\!_{ij}$.
Meanwhile the scalar wave equation becomes
 \be
0=\rho\phi''+\left(\alpha+1-\frac{d}2\right)\phi'+\frac12
\partial_\rho\log(-g)\left(\frac{\alpha}2\phi+\rho\phi'\right)+\frac{L^2}4\Box_g\phi
-\frac{\kappa}8
\rho^{\frac{\alpha}2-1}\phi^2+O(\rho^{\alpha-1}\phi^3)\,.
 \labell{scala1}
 \ee
Here we eliminated the leading term in this equation since vanishes for
$\alpha=d-\Delta$ or $\Delta$, just as in the probe analysis leading to
eq.~\reef{field0}.  Further, we have defined
$\Box_g\phi\equiv\frac1{\sqrt{-g}}\partial_i\left(
\sqrt{-g}g^{ij}\partial_j\phi\right)$ where the full metric
$g_{ij}(x,\rho)$ is again inserted in this wave operator along the
boundary directions.

Now we are in a position to construct solutions with a small-$\rho$
expansion near the asymptotic boundary. We should say that our
objective here is primarily to understand the form of these solutions.
We will delay considering explicit solutions to the next two sections.
Hence, to get a feeling for the expansion of $g_{ij}(x,\rho)$, we
consider the equations of motion \reef{einsteinx} and \reef{scala1}
with
 \be
g_{ij}(x,\rho)\simeq\,\gz_{ij}(x)+\rho^\beta \gf{\beta}_{ij}(x) \qquad
{\rm and}\qquad \phi(x,\rho)\simeq\phi^{(0)}(x)+\rho^{\beta'}
\phi^{(\beta')}(x)\,,
 \labell{testx}
 \ee
where we assume both exponents $\beta$, $\beta'$ are positive. We begin
by examining the first equation in eq.~\reef{einsteinx} to linear order
in $\overset{\scriptscriptstyle{(\beta)}}{g}\!_{ij}$. First we find
that there are two homogeneous solutions, \ie solving with all of the
terms linear in $g''$ and $g'$. That is, we have either $\beta=0$ or
$\beta=d/2$ where the latter also requires
$\overset{\scriptscriptstyle{(0)}}{g}\,\!^{ij} \,
\overset{\scriptscriptstyle{(\beta)}}{g}\!_{ij}=0$. The first case is
simply deformation of the boundary metric
$\overset{\scriptscriptstyle{(0)}}{g}_{ij}$ where as the second is the
second independent solution containing the state data about the
stress-energy, as in eq.~\reef{stress}. This structure matches
precisely that found for the usual FG expansion \reef{expand} and
depends only on the asymptotic geometry approaching AdS geometry. At
this linear level, $R_{ij}(\overset{\scriptscriptstyle{(0)}}{g})$
introduces an inhomogeneous source term requiring $\beta=1$. Similarly
$\phi^{(0)}$ introduces various source terms on the right-hand side.
The leading source comes from the mass term which requires
$\beta=\alpha$, while the next source would be the cubic term which
calls for $\beta=3/2\,\alpha$. Hence in the deformed background, the
asymptotic expansion of $g_{ij}(x,\rho)$ involves terms with two powers
of $\rho$, namely, integer powers $\rho^n$ as well as powers
$\rho^{m\alpha/2}$. To simplify the general expansion in a workable
form, we consider the case where $\alpha/2$ is a rational number, \ie
$\alpha/2=N/M$ where $N$ and $M$ are relatively prime. In this case,
the general asymptotic expansion for the metric $g_{ij}(x,\rho)$ can be
written as
 \bea
g_{ij}(x,\rho)&=&\sum_{n=0}^{N-1}\left(\rho^n\,\gf{n}_{ij}(x)+\sum_{m=2}^\infty
\rho^{n+m\frac{\alpha}2}\,\gf{n+m\frac{\alpha}2}_{ij}(x)\right)
 \labell{expandgx}\\
&&\qquad+\ \rho^{d/2}\,\sum_{n=0}^{N-1}\left(\rho^n\,
\gf{\frac{d}2+n}_{ij}(x) +\sum_{m=2}^\infty \rho^{n+m\frac{\alpha}2}\,
\gf{\frac{d}2+n+m\frac{\alpha}2}_{ij}(x)\right)\,.\nonumber
 \eea
If $\rho^{d/2}$ appears in the series in the first line, the expansion
contains a logarithmic term
 \bea
g_{ij}(x,\rho)&=&\sum_{n=0}^{N-1}\left(\rho^n\,\gf{n}_{ij}(x)+\sum_{m=2}^\infty
\rho^{n+m\frac{\alpha}2}\,\gf{n+m\frac{\alpha}2}_{ij}(x)\right)
 \labell{expandgx2}\\
&&\qquad+\ \rho^{d/2}\log\rho\,\sum_{n=0}^{N-1}\left(\rho^n\,
\fff{\frac{d}2+n}_{ij}(x) +\sum_{m=2}^\infty \rho^{n+m\frac{\alpha}2}\,
\fff{\frac{d}2+n+m\frac{\alpha}2}_{ij}(x)\right)\,.\nonumber
 \eea
Of course, the latter expansion with the logarithmic contribution
always arises if $d$ is even, as in the usual FG expansion
\reef{expand}. However, we note that this form can also arise in odd
dimensions if the relevant operator has an appropriate dimension. For
example, if $d=3$, eq.~\reef{expandgx2} arises for $\alpha=3/m$ with
$m=2, 3, 4,..,$ which would correspond to $\Delta=3(m-1)/m$. In
particular here, $m=3$ yields $\Delta=2$ which corresponds to the
fermion mass term in $d=3$.

We might also note some simplifications that can occur in the above
expansions. In particular, if the boundary curvature vanishes, the
integer powers are not required, \ie all of the coefficients with $n>0$
vanish --- see the explicit solutions in section \ref{exercise}.
Similarly for the case of a free scalar in the bulk or a scalar theory
that is invariant under $\Phi\to-\Phi$, only (integer) powers of
$\rho^\alpha$ will appear, \ie all of the coefficients with $m$ being a
odd integer vanish.

Now using the test expansion \reef{testx}, we can also examine the
scalar wave equation \reef{scala1} to linear order in
$\phi^{(\beta')}$. Considering only the first two linear terms, we find
two homogeneous solutions, namely $\beta'=0$ and $\beta'=d/2-\alpha$.
The first case is simply a shift of the boundary coupling $\phi^{(0)}$,
whereas the second power corresponds to the second independent
solution. The overall power of this contribution is then
$\rho^{\frac{\alpha}2+\beta'}=\rho^{\frac{d-\alpha}2}=\rho^{\Delta_+/2}$.
Hence we see again that this structure matches precisely that found
with the probe analysis in eq.~\reef{field0}, which depends only on the
fact that the asymptotic geometry approaches the AdS geometry. Further,
as before, the second solution contains the expectation value of the
corresponding boundary theory operator, as in eq.~\reef{expect2}.
Continuing with the linearized analysis, $\phi^{(0)}$ introduces two
inhomogeneous source terms in eq.~\reef{scala1} from the derivative
term $\Box_g\phi$ and from the higher order contributions from the
scalar potential. The former requires $\beta'=1$ while the latter
requires $\beta'=\alpha/2$. Hence in the full nonlinear solution, we
see that the asymptotic expansion of $\phi(x,\rho)$ also involves two
powers of $\rho$, namely integer powers of $\rho$ and $\rho^{\alpha/2}$
separately, as in the metric expansion above. Given the expansions
\reef{expandgx} and \reef{expandgx2}, we see the metric also feeds in
source terms with both kinds of powers from the term with
$\partial_\rho\log(-g)$. If we simplify the general expansion with the
choice $\alpha/2=N/M$ as above, we find
 \beqa
\Phi(x,\rho)=\rho^{\alpha/2}\phi(x,\rho)&=& \rho^{\alpha/2}
\sum_{n=0}^{N-1}\,\sum_{m=0}^\infty
\rho^{n+m\frac{\alpha}2}\,\phi^{(n+m\frac\alpha2)}(x)
 \labell{expand2}\\
 &&\qquad +
\rho^{(d-\alpha)/2} \sum_{n=0}^{N-1}\,\sum_{m=0}^\infty
\rho^{n+m\frac{\alpha}2}\,\phi^{(\frac{d}2-\alpha+n+m\frac\alpha2)}(x)\,.
 \nonumber
 \eeqa
In this case when $d$ is even or when $d$ is odd and
$\alpha=(d-2n)/(m+2)$ (subject to the condition $\al>0$), the powers in
the second series actually overlap with those in the first. Hence in
this case, the second independent solution actually has an extra factor
of $\log\rho$, which gives rise to the following expansion
 \beqa
\Phi(x,\rho)=\rho^{\alpha/2}\phi(x,\rho)&=& \rho^{\alpha/2}
\sum_{n=0}^{N-1}\,\sum_{m=0}^\infty
\rho^{n+m\frac{\alpha}2}\,\phi^{(n+m\frac\alpha2)}(x)
 \labell{expand3}\\
 &&\qquad +
\rho^{(d-\alpha)/2}\log\rho \sum_{n=0}^{N-1}\,\sum_{m=0}^\infty
\rho^{n+m\frac{\alpha}2}\,\psi^{(\frac{d}2-\alpha+n+m\frac\alpha2)}(x)\,.
 \nonumber
 \eeqa
Further the leading coefficient $\psi^{(\frac{d}2-\alpha)}(x)$ can be
related to matter conformal anomalies \cite{construct,anom}.

Now before leaving our discussion of the back-reacted bulk solution, we
wish to comment on the fixed boundary data. While the details of the
small-$\rho$ expansion in the metric \reef{back0} have changed, the
second independent solution still arises at order $\rho^{d/2}$ with the
coefficient $\overset{\scriptscriptstyle{(d/2)}}{g}\!_{ij}$. As before,
this coefficient carries information about the state of the boundary
field theory through the relation in eq.~\reef{stress}. Similarly, the
second independent solution appears in the expansion of the scalar
field at order $\rho^{(d-\alpha)/2}$, which again is determined by the
state of the boundary theory through eq.~\reef{expect2}. Hence we may
ask at what order the state data for the scalar, \ie
$\phi^{(\frac{d}2-\alpha)}$, begins to contribute to the expansion of
the metric. Examining the Einstein equation \reef{einsteinx}, we see
the leading contribution will come from the mass term on the right-hand
side with a cross term $\phi^{(0)}\phi^{(\frac{d}2-\alpha)}$. However,
this contribution enters with a factor $\rho^{d/2-1}$ and so we can see
that $\phi^{(\frac{d}2-\alpha)}$ will only effect the coefficients in
the metric expansion $\overset{\scriptscriptstyle{(n)}}{g}\!_{ij}$ with
$n\ge d/2$. Hence we can still refer to the metric coefficients with
$n<d/2$ as the fixed boundary data, as well as $\phi^{(n)}$ with $n<d/2
- \alpha$, since these coefficients are all independent of the state of
the boundary theory and are completely fixed by the boundary metric
$\overset{\scriptscriptstyle{(0)}}{g}\!_{ij}$ and the coupling
$\phi^{(0)}$.

We note that the calculation of the holographic EE is a geometric one
which relies on the details of the metric expansion \reef{expandgx} or
\reef{expandgx2}, \ie the scalar field does not directly enter into the
extremal area \reef{define}. Hence, as in the previous section, we
would like to show that the logarithmic contribution to the holographic
EE only depends on the fixed boundary data, \ie
$\overset{\scriptscriptstyle{(n)}}{g}\!_{ij}$ with $n<d/2$. Hence, we
must next examine the embedding functions $X^\mu(y^a,\tau)$ to show
that this universal contribution also only depends on the geometry of
the entangling surface in the boundary and is independent of the
details of the bulk surface, \eg ensuring that it has a regular
geometry.

The embedding functions are determined by extremizing the area of the
bulk surface $\m$. It is a straightforward exercise to show that this
leads to the following (local) equation of motion
 \be
\frac{1}{\sqrt{h}}\partial_\alpha\left(\sqrt{h}h^{\alpha\beta}
\partial_\beta X^\mu\right)+h^{\alpha\beta}\Gamma^\mu{}_{\nu\sigma}
\partial_\alpha X^\nu\partial_\beta X^\sigma =0\,,
 \labell{eomx}
 \ee
where the induced metric is given by eq.~\reef{induceh} and
$\Gamma^\mu{}_{\nu\sigma}$ denote the usual Christoffel symbols
constructed with the bulk metric $g_{\mu\nu}$. As before, we make the
gauge choices given in eq.~\reef{eqn:gauge-brane} and in this case,
setting $\mu=\rho=\tau$ in eq.~\reef{eomx} yields a spurious
constraint, which is automatically satisfied upon solving the remaining
equations for $X^i(y^a,\tau)$. Of course, the leading terms in a
small-$\tau$ expansion of the latter are just the constant terms
describing the position of the entangling surface in the boundary, \ie
$X^i(y^a,\tau) \simeq \overset{\scriptscriptstyle{(0)}}{X^i}(y^a)$. To
determine the order at which a second independent solution appears, we
follow the analysis in \cite{grwit}. We begin by assuming the equations
\reef{eomx} have been solved perturbatively to $o(\tau^s)$, such that
the coefficients $\overset{\scriptscriptstyle{(s)}}{X^i}(y^a)$ are to
be solved in terms of the previous terms. One can see that the leading
contribution of these new coefficients always comes from the terms in
eq.~\reef{eomx} with two $\tau$ derivatives. If we substitute in the
leading form of the induced metric \reef{inducemet}, this term is given
by
 \bea
0&\simeq&\frac{4}{L^2}\frac{\tau^{d/2}}{\sqrt{\hz}}\,\partial_\tau\!
\left( \sqrt{\hz} {\tau^{1-\frac{d}{2}}}\, \partial_\tau\! \left(
\tau^{s} \overset{\scriptscriptstyle{(s)}}{X^i}(y^a)\right)\right) +
\frac{4\tau^2}{L^2}\, g^{ik}\partial_\tau g_{kj}\, \partial_\tau\!
\left( \tau^{s} \overset{\scriptscriptstyle{(s)}}{X^j}(y^a)\right)
+\cdots
 \nonumber\\
&\simeq& s\left(s-\frac{d}{2}\right)\tau^{s-1}
\overset{\scriptscriptstyle{(s)}}{X^i}(y^a)+ \cdots\,,
 \labell{undeter}
 \eea
Now the latter result implies that
$\overset{\scriptscriptstyle{(s)}}{X^i}$ becomes undetermined for
$s=d/2$ and it can be independently specified, \eg to ensure that the
extremal surface has a regular geometry. Note that this is precisely
the same order at which this additional information entered in the
previous section (without the relevant deformation). Further, this is
also precisely the order at which the second independent set of
coefficients appear in the small-$\tau$ expansion of the bulk metric.
If we examine the full equations of motion \reef{eomx} for the
embedding functions in more detail, we also find that, just as in the
expansions for the metric and the scalar, there are two powers of
$\tau\ (=\rho)$ appear, namely, powers of $\tau$ and $\tau^{\alpha/2}$.
Further, it is straightforward to show that the small-$\tau$ expansion
for $X^i(y, \tau)$ takes an analogous form as that presented for the
bulk metric in eq.~\reef{expandgx} (or eq.~\reef{expandgx2}, depending
on the precise values of $d$ and $\alpha$). Of course, the leading
coefficients $\overset{\scriptscriptstyle{(n)}}{X^i}$ with $n<d/2$ are
completely determined as local functionals of
$\overset{\scriptscriptstyle{(0)}}{X^i}(y^a)$,
$\overset{\scriptscriptstyle{(0)}}{g}_{ij}$ and $\phi^{(0)}$.

Combining the asymptotic boundary expansions for the bulk metric and
the embedding functions, one produces a similar expansion for the
induced metric \reef{induceh}, \eg
 \bea
h_{ab}(y,\tau)&=&\frac{1}{\tau}\left[\,\sum_{n=0}^{N-1}\left(\tau^n\,
\hf{n}_{\,ab}(y)+\sum_{m=2}^\infty
\tau^{n+m\frac{\alpha}2}\,\hf{n+m\frac{\alpha}2}_{ab}(y)\right)\right.
 \labell{expandhx}\\
&&\left.\qquad+\ \tau^{d/2}\,\sum_{n=0}^{N-1}\left(\tau^n\,
\hf{\frac{d}2+n}_{ab}(y) +\sum_{m=2}^\infty \tau^{n+m\frac{\alpha}2}\,
\hf{\frac{d}2+n+m\frac{\alpha}2}_{ab}(y)\right)\right]\,.\nonumber
 \eea
Of course, if $\tau^{d/2}$ appears in the series in the first line
above, then a logarithmic factor will appear in the second line, as in
eq.~\reef{expandgx2}. Recall that the leading coefficient
$\overset{\scriptscriptstyle{(0)}}{h}_{ab}$ is the induced metric
$H_{ab}$ on the entangling surface $\Sig$ in the background for the
boundary theory. The component $h_{\tau\tau}(y,\tau)$ has an analogous
expansion with a pre-factor $L^2/(4\tau^2)$ and
$\overset{\scriptscriptstyle{(0)}}{h}_{\tau\tau}=1$, as in
eq.~\reef{inducemet}. For the present purposes, an essential feature of
the induced metric is that all of the coefficients
$\overset{\scriptscriptstyle{(n)}}{h}_{\alpha\beta}$ depend only on the
fixed boundary data for $n<d/2$. That is, these leading coefficients
are again completely determined by
$\overset{\scriptscriptstyle{(0)}}{X^i}(y^a)$,
$\overset{\scriptscriptstyle{(0)}}{g}_{ij}$ and $\phi^{(0)}$.

Now turning to the holographic EE \reef{define}, we must evaluate the
area of the extremal surface. The area integral has the same basic
structure as given in eq.~\reef{arealx} in the absence of a relevant
deformation. In particular, the leading expression provides a factor of
$\tau^{-d/2}$ and the radial integral ends at the regulator surface
with $\tau_{min}=\delta^2/L^2$ where $\delta$ is the UV cut-off in the
boundary theory. We are again primarily interested in the contribution
proportional to $\log\delta$ and so we must expand the rest of the
integrand to order $\tau^{\frac{d-2}2}$. While the details of this
expansion are now modified by the presence of the relevant deformation,
as before, it suffices to observe that a term at the desired order will
only contain the coefficients
$\overset{\scriptscriptstyle{(n)}}{h}_{\al\bt}$ with $n\le(d-2)/2$.
Hence this logarithmic contribution to the holographic EE is completely
determined by the fixed boundary data. That is, this contribution is
completely determined by $\overset{\scriptscriptstyle{(0)}}{X^i}(y^a)$,
$\overset{\scriptscriptstyle{(0)}}{g}_{ij}$ and $\phi^{(0)}$. In fact,
the same result also applies for all of the divergent contributions to
the holographic EE.

While this conclusion has been unchanged by the introduction of a
relevant operator in the boundary theory, the appearance (or not) of a
universal contribution in the holographic EE, proportional to
$\log\delta$, depends very much on the details, \ie on the dimension of
the operator, as well as the spacetime dimension. In particular, in the
expansion of the integrand in eq.~\reef{arealx}, we must identify a
higher order term with $\tau^{n+m\frac{\alpha}{2}}=\tau^{\frac{d-2}2}$.
Of course, as in the previous section, one finds such terms with $m=0$
for any even $d$. However, there can now be new terms for odd or even
$d$ if the operator dimension of the relevant deformation is
appropriate. For example, choosing an operator with
$\Delta=\frac{d}2+1$ yields $\alpha=(d-2)/2$ and hence we find the
desired power of $\tau$ with $n=0$ and $m=2$. Similarly, for a scalar
mass term with $\Delta=d-2$, one finds $\alpha=d-\Delta=2$ and hence a
logarithmic term appears in even dimensions with $d\ge
6$.\footnote{Recall that $m\ge2$ (or $m=0$) because the stress tensor
for the Einstein-scalar theory in the bulk is at least quadratic in the
scalar field, \eg all of the contributions on the right-hand side of
eq.~\reef{einsteinx} are quadratic or higher order in the scalar.} One
can compare this to the results of \cite{frank}, where an analogous
contribution \reef{franks} was found for a free massive scalar field.
As a final note here, we observe that in certain instances (with
appropriate $\alpha$ and $d$) the logarithmic term will be produced
with both $n$ and $m$ nonvanishing. In the following sections, we
present some explicit calculations of these new universal contributions
to the holographic EE.

\section{New Universal Terms with a Deformed Boundary Theory} \label{exercise}

While our general discussion above indicated that a relevant
deformation of the boundary theory may lead to new universal
contributions in the holographic EE, we would like to present some
explicit examples where such logarithmic contributions appear. To make
the problem of explicitly calculating the new universal terms simpler,
we begin by considering here a background with a flat boundary metric.
For convenience, we also depart from the conventions of the previous
section by choosing a new radial coordinate $z$ where $\rho=z^2/L^2$.
We take the Einstein-scalar theory in eq.~\reef{action} with
 \beq V(\Phi)=-\frac{d(d-1)}{L^2}+\frac12
m^2\Phi^2+\frac{\kappa}{6L^2}\Phi^3\,,
 \labell{pot2}
 \eeq
\ie we choose the potential to include only terms up to cubic in the
scalar. Now with a flat boundary, our asymptotically AdS$_{d+1}$ bulk
metric becomes
 \be
ds^2=\frac{L^2}{z^2}\left(dz^2+f(z)\, \eta_{ij} dx^i dx^j\right)\,,
 \labell{back2}
 \ee
and following eq.~\reef{scala}, we write the scalar profile as
$\Phi(z)=(z/L)^\al\phi(z)$. Asymptotically, as $z\to 0$, $f(z)\to1$ and
$\phi(z)\to\phi^{(0)}$.

Examining the Einstein equations of motion \reef{Einstein}, there are
two nontrivial equations which may be written:
 \beqa
\frac{d(d-1)}2\left[\left(\frac{f'}{f}\right)^2-\frac4z\frac{f'}f\right]
-\Phi'^2+\frac{(mL)^2}{z^2}\Phi^2+\frac{\kappa}{3z^2}\Phi^3&=&0\,,\labell{zz}\\
2(d-1)\left[f''-\frac{d-1}z f'+\frac{d-4}4\frac{f'^2}{f}\right]
+f\left(\Phi'^2+\frac{(mL)^2}{z^2}\Phi^2+\frac{\kappa}{3z^2}\Phi^3\right)&=&0\,.\labell{tt}
 \eeqa
However, the Bianchi identity ensures that these equations (combined
with the scalar field equation) are redundant. For simplicity we focus
on eq.~\reef{zz} in the following. The scalar field equation
\reef{scalar} reduces to
 \be
\Phi'' -\frac{d-1}z \Phi' +\frac{d}2\frac{f'}f
\Phi'-\frac{(mL)^2}{z^2}\Phi-\frac{\kappa}{2z^2}\Phi^2=0\,.
 \labell{scalar2}
 \ee

Now constructing power series solutions for $f$ and $\phi$ around
$z=0$, one finds
 \beqa
  f(z)&=&1+\sum_{k=2} a_k \left(\phi^{(0)}\,(z/L)^\alpha\right)^{k}\,,
\labell{series}\\
\phi(z)&=&\phi^{(0)}\,(z/L)^\alpha+\sum_{k=2} b_k
\left(\phi^{(0)}\,(z/L)^\alpha\right)^{k}\,.
 \nonumber
 \eeqa
Note that we are being somewhat cavalier in both of these expansions
since neither includes the second independent solution shown in, \eg
eqs.~\reef{expandgx} and \reef{expand2}. However, as we showed in the
previous section, none of the terms which have been neglected will
contribute to the logarithmic contributions in the holographic EE. In
comparing the above expression for $f(z)$ with eq.~\reef{expandgx}, we
see that the terms involving $n\ne0$ do not appear above. This
simplification occurs because the boundary curvature vanishes.

The first few coefficients in the above expansions are explicitly
determined to be
 \beqa
a_2&=& -\frac1{4 (d-1)}\,,\qquad
b_2=-\frac{\kappa}{2\alpha(d-3\alpha)}\,,\labell{coeffs}\\
a_3&=&\frac{2\kappa}{9\alpha(d-1)(d-3\alpha)}\,,
 \nonumber\\
b_3&=&-\frac{d\,\alpha}{8(d-1)(d-4\alpha)}+\frac{\kappa^2}{
4\alpha^2(d-3\alpha)(d-4\alpha)}\,,
 \nonumber\\
a_4&=&\frac{(3d-8)\alpha+2d }{64(d-1)^2(d-4\alpha)} -
\frac{\kappa^2(5d-17\alpha)}{32\alpha^2(d-1)(d-3\alpha)^2(d-4\alpha)}\,,
 \nonumber\\
b_4&=&\frac{\kappa\, d (17d-65\alpha) }{ 72(d-1)(d-3\alpha)(d-4\alpha)
(d-5\alpha)} -
\frac{\kappa^3(3d-10\alpha)}{24\alpha^3(d-3\alpha)^2(d-4\alpha)(d-5\alpha)}\,,
 \nonumber\\
a_5&=&\frac{\kappa\,((79d-200)\alpha^2-(19d-90)d\alpha -10 d^2)}{180
\alpha(d-1)^2(d-3\alpha)(d-4\alpha)(d-5\alpha)}
+\frac{\kappa^3(3d-13\alpha)}{30\alpha^3(d-1)(d-3\alpha)^2(d-4\alpha)(d-5\alpha)}\,,
 \nonumber\\
b_5&=&\frac{3d^2\alpha^2}{64(d-1)^2(d-4\alpha)(d-6\alpha)} -
\frac{\kappa^2d(2655\alpha^2-1330\alpha d+163d^2)
}{576\alpha(d-1)(d-3\alpha)^2(d-4\alpha)(d-5\alpha)(d-6\alpha)}
 \nonumber\\
&&\qquad\qquad+
\frac{\kappa^4(6d-25\alpha)}{96\alpha^4(d-3\alpha)^2(d-4\alpha)
(d-5\alpha)(d-6\alpha)}
 \,.\nonumber
 \eeqa
Here, we have used eqs.~\reef{zz} and \reef{scalar2} to determine these
coefficients. As an extra check, we also explicitly checked that the
above coefficients also solve eq.~\reef{tt} to order
$\left(\phi^{(0)}\,(z/L)^\alpha\right)^5$. Note that if we set
$\kappa=0$, the only nonvanishing coefficients are $a_k$ with even $k$
and $b_k$ with odd $k$.

Recall that the calculation of the holographic EE \reef{define} is
purely a geometric one and the scalar field only effects the result
through its back-reaction on the bulk metric. Hence to evaluate
eq.~\reef{define}, we need only focus on the expansion of the metric,
\ie the expansion of $f(z)$ in powers of $z^\alpha$, as seen in
eq.~\reef{series}. As shown in section \ref{rest}, the universal part
of holographic EE involves only the terms in the expansion up to the
power just proceeding $z^d\ (\simeq \rho^{d/2})$. Hence the expansion
\reef{series} of $f(z)$ must be carried out to a maximum value of $k$:
 \be
d-\al\le \alpha\, k_{max}< d\,.
 \labell{kmax}
 \ee
Recall from eq.~\reef{range0}, we also have $0\le\al\le d/2$. Combining
this inequality \reef{kmax} with the above expansion \reef{series}, we
see that $k_{max}=2,3,4$ or 5 --- for which we can read the
coefficients from eq.~\reef{coeffs} --- is sufficient for
$d/3\leq\al\leq d/2\, ,~d/4\leq\al\leq d/3$, $d/5\leq\al\leq d/4$ or
$d/6\leq\al\leq d/5$, respectively. In general, a given $k_{max}$ is
sufficient for $d/(1+k_{max})\leq\al\leq d/k_{max}$. To proceed
further, we choose explicit values of $d$ and $\Delta$, as well as the
geometry of the entangling surface. In particular for the latter, we
consider 1) two flat planes bounding a slab and 2) a spherical surface
$S^{d-2}$.

Before proceeding with explicit calculations, we address a question
about interpreting the results in terms of the boundary theory. Note
that with the present conventions, $\phi^{(0)}$ is a dimensionless
parameter. However, following the standard AdS/CFT dictionary, we wish
to relate this parameter in terms of the coupling to an operator with
conformal dimension $\Delta$ in the boundary theory. As such, this
coupling should be dimensionful defining some mass scale with
$\phi^{(0)}\sim\mu^{d-\Delta}$. The question is then how to make this
relation more precise, \eg what scale enters on the bulk side of this
equation? Of course, in the AdS/CFT correspondence, the natural scale
emerging from the bulk theory is simply the AdS curvature scale
yielding
 \be
\frac{\phi^{(0)}}{L^{d-\Delta}} = \lambda\,\mu^{d-\Delta}\,.
 \labell{convertp}
 \ee
Here it is convenient to introduce a dimensionless parameter $\lambda$,
which would control the strength of the deformation in the boundary
theory.\footnote{Upon converting our results below to parameters of the
boundary theory, the power of $\lambda$ keeps track of the number
factors of $\phi^{(0)}$ appearing in the holographic calculation.} Note
that given the present framework, we can not provide a more specific
relation than eq.~\reef{convertp} above. For example, if we consider a
mass deformation like $m^2 \phi^2$ in the boundary field theory, we
could always redefine the operator by numerical factors, \eg the dual
operator could equally well be $\phi^2$ or $\phi^2/2$ or $\sqrt{2} \pi
\phi^2$ and then accordingly the coupling would be $m^2$ or $2 m^2$ or
$m^2/(\sqrt{2} \pi)$. This example illustrates that distinguishing the
operator from the coupling part will not be well-defined without some
additional information about the boundary theory. In fact, in certain
cases, the required information may be provided by supersymmetry and
knowing more details of the duality between the bulk and boundary
theories. One such example would be $N=2^*$ theories \cite{nstar2},
where more precise results can be obtained \cite{psi}.

\subsection{Flat entangling surfaces}
\label{sec:belt}

In this case, we introduce two flat parallel planes as the entangling
surface. Hence subsystem of interest in the boundary theory is the
following slab: $V_\mt{F}=\{0\le x^1\le \ell,\ t=0\}$. The holographic
EE has been calculated for this geometry in the case where the boundary
theory is simply a $d$-dimensional CFT \cite{rt1,rt2}:
 \be
S_\mt{CFT}(V_\mt{F})=\frac{4\pi}{d-2}\frac{L^{d-1}}{\lp^{d-1}}\left[
\frac{R^{d-2}}{\delta^{d-2}} - \gamma_d\, \frac{R^{d-2}}{\ell^{d-2}}
\right]\,,
 \labell{olda2}
 \ee
where $\gamma_d$ is a numerical factor: $\gamma_d=\frac12
\left(2\sqrt{\pi}\,\Gamma\!\big(\frac{d}{2(d-1)}
\big)/\,\Gamma\!\big(\frac{1}{2(d-1)} \big) \right)^{\!d-1}$. Here $R$
is a regulator length along the $x^{2,3,\cdots,d-1}$ directions, which
was introduced so that the entangling surfaces have a finite area, \ie
$R^{d-2}$. Hence for a CFT in this geometry, no logarithmic
contribution appears in the EE for either even or odd $d$. Note that
the pre-factor in the above expression can be interpreted as a central
charge in the boundary theory, \eg the leading singularity in the
two-point of the stress tensor is controlled by the central charge, \ie
$C_T\simeq L^{d-1}/\lp^{d-1}$.

Returning to the holographic EE in the presence of a relevant
deformation, we describe the bulk surface $\m$ with the profile
$z=z(x^1)$ with the boundary conditions $z(x^1=0) =0= z(x^1=\ell)$. The
induced metric $h_{\al\bt}$ on this surface embedded into the
background (\ref{back2}) is given by
 \be
h_{\al\bt}\,dx^{\al}dx^{\bt}=\frac{L^2}{z^{2}}\left[(f(z)+\dot
z^2)\,(dx^1)^2 + f(z)\sum_{i=2}^{d-2}(dx^i)^2 \right]
 \quad,
 \ee
where `dot' denotes a derivative with respect to $x^1$. Evaluating the
area of $\m$ in the bulk then yields
 \be
 A(\m)=\int\prod_{i=1}^{d-1}dx^i\sqrt{\mathrm{det}h_{\al\bt}}
 =R^{d-2}L^{d-1}\int_{0}^{\ell} dx^1 \frac{f^{\,d/2-1}}{z^{\, d-1}}\sqrt{f+\dot z^2}
 \, .
 \ee
We treat this expression as an action for $z(x^1)$ to determine the
profile which extremizes this area. As the action contains no explicit
$x^1$ dependence, the conjugate `energy' is conserved. This conserved
energy functional then yields the following equation:
 \be
 \dot z^2 = \Big(\frac{z_*}{z}\Big)^{2(d-1)}\frac{f^{\,d}}{f_*^{\,d-1}}-f
 \,,
 \ee
where $z_*$ corresponds to the (maximum) value of $z$ where $\dot z =0$
and $f_*=f(z_*)$. From the inversion symmetry, $x^1\rightarrow
\ell-x^1$, we must have $\dot z =0$ at $x_1=\ell/2$. To determine
$z_*$, we can integrate the above equation
 \be
 \frac{\ell}{2}=\int_0^{z_*}dz\, \Big(\frac{z}{z_*}\Big)^{d-1}
 \( {f^{\,d-1}\over f_*^{\,d-1}}-\Big({z \over z_*}\Big)^{2(d-1)} \)^{-1/2}f^{-1/2}
 \, .
 \ee

With these results, we can evaluate the holographic EE as follows
 \be
S(V_\mt{F})={2\pi \over \lp^{d-1}}A(\m) =4\pi {L^{d-1} \over
\lp^{d-1}}\,R^{d-2}\int_{\delta}^{z_*}  dz\, {f^{\,d/2-1}  \over
z^{d-1}}
\[ 1-\Big({f_* \, z^2\over f\,z_*^2 }\Big)^{\,d-1}  \]
 \,,
 \labell{EEbelt}
 \ee
where we have introduced the UV regulator surface at $z=\delta$. We are
interested in extracting a universal (logarithmic) contribution from
the above expression, in the limit $\delta\rightarrow 0$. Therefore we
expand the integrand in powers of $z$ and evaluate only the term with
$1/z$. In fact, the expression within the square brackets can be set to
1, since the higher order terms which it contributes in this expansion
begin at $z^{d-1}$. Therefore the desired universal coefficient is
independent of $z_*$. In the present notation, $z_*$ represents the
undetermined data, appearing at higher order in the embedding
functions, which is specified to produce a smooth surface $\m$ in the
bulk. Hence, as discussed around eq.~\reef{undeter}, this data will not
contribute to the universal terms in the holographic EE. Further, this
observation allows us to consider the limit $\ell\to\infty$ in which
case $z_*\to\infty$ and the expression inside the square brackets above
simply reduces to 1. In this limit, we are simply calculating the
entanglement entropy upon dividing the boundary theory into two
(semi-infinite) regions with a single wall at $x^1=0$. In the case
where the boundary theory is conformal, this limit leaves only the
regulator dependent term, as shown in eq.~\reef{olda2}. However, the
limit leaves a more interesting result in the present case because the
relevant deformation in the boundary theory has introduced a finite
correlation length, $\xi=1/\mu$. In particular, as we now show
explicitly, the result can include a universal logarithmic
contribution, of the form found in \cite{frank}.

As noted above, a logarithmic contribution will only appear from a term
in the small-$z$ expansion of the factor $f^{\,d/2-1}$ in
eq.~\reef{EEbelt}. Further, we note that given the form \reef{series}
of $f$, this expansion only produce powers $z^{m\alpha}$ with $m\ge2$.
Hence to produce a $1/z$ term in the integrand, we must have
$\alpha=(d-2)/m$. Note that all such values appear in the allowed range
given in eq.~\reef{range0} and the conformal dimension of the dual
operator would be $\Delta=d-\frac{d-2}m$. We consider explicit examples
for specific values of $m$ below.

\subsubsection{$m=2$} \label{eme2a}

In this case, we have $\al=(d-2)/2$ and we only need the first term in
the expansion (\ref{series}) of $f$. Examining eq.~(\ref{EEbelt}), we
find
 \be
S(V_\mt{F})={\pi \over 2}{d-2 \over d-1} {L^{d-1} \over
\lp^{d-1}}\,\lambda^2 R^{d-2}\mu^{d-2}\,\log \mu\delta + \cdots
 \labell{EEbelt1}
 \ee
which applies for any odd or even $d\ge3$. Above, we have used
eq.~\reef{convertp} to write $(\phi^{(0)})^2/L^{d-2} = \lambda^2
\mu^{d-2}$. We have also introduced a factor of $\mu$ to make the
argument of the logarithm dimensionless, since it is the only natural
scale to appear there. We are implicitly assuming an operator arises
with a specific conformal dimension which is dependent on $d$. However,
we might note that for $d=4$, $\Delta=3$ which corresponds to a fermion
mass term while for $d=6$, $\Delta=4$ which corresponds to a scalar
mass term.

\subsubsection{$m=3$} \label{firtree}

In this case, $\al=(d-2)/3$ and we must expand eq.~(\ref{series}) to
order $k_{max}=3$. Eq.~\reef{EEbelt} then yields
 \be
S(V_\mt{F})=-{ 2\, \pi \kappa \over 3(d-1)}  {L^{d-1} \over
\lp^{d-1}}\,\lambda^3 R^{d-2}\mu^{d-2}\, \log \mu\delta + \cdots
 \labell{EEbelt1a}
 \ee
using $(\phi^{(0)})^3/L^{d-2} = \lambda^3 \mu^{d-2}$. This result again
applies for any odd or even $d\ge3$. Note that this contribution
vanishes for $\kappa=0$, \eg for a free bulk scalar.

\subsubsection{$m=4$}  \label{firtree2}

With $m=4$, $\al=(d-2)/4$ and we expand eq.~(\ref{series}) to order
$k_{max}=4$. Then from eq.~\reef{EEbelt}, we obtain
 \be
S(V_\mt{F})=\[{ 2(3d+34) \over (d-2)(d+6)^2 }\kappa^2-{(3d+8)(d-2)^2
\over 256 (d-1)}  \] {\pi \over (d-1)} {L^{d-1} \over
\lp^{d-1}}\,\lambda^4 R^{d-2}\mu^{d-2}\,\log\mu \delta + \cdots
 \labell{EEbelt2}
 \ee
using $(\phi^{(0)})^4/L^{d-2} = \lambda^4 \mu^{d-2}$. Again, we are
implicitly assuming a specific operator dimension which is dependent on
$d$ but with this assumption, the corresponding universal contribution
will appear for any odd or even $d\ge3$.

\subsection{Spherical entangling surfaces}
\label{sec:disk}

In this case, we wish to calculate the EE across a spherical surface in
the boundary theory. If we define the radial coordinate as usual, \ie
$r^2= \sum_i (x^i)^2$, in the flat boundary geometry, then the relevant
subsystem is the ball: $V_\mt{S}=\{r\le R,\ t=0\}$. Again, the
holographic EE has been calculated in this case with a conformal
boundary theory and a logarithmic contribution arises for even $d$
\cite{rt1,rt2}:
 \be
S_\mt{CFT}(V_\mt{S})=(-)^{\frac{d}2-1}\frac{4\pi^{d/2}}{\Gamma(d/2)}
\frac{L^{d-1}}{\lp^{d-1}}\log\left(2R/\delta\right)+\cdots\,.
 \labell{oldb3}
 \ee
In fact, this result can be calculated for any CFT without any
reference to holography and it is known that the pre-factor is
precisely $(-)^{\frac{d}2-1}4A$ \cite{solo,new,cthem2,circle4} where
the $A$ is the central charge appearing in the $A$-type trace anomaly
\cite{deser}. This contribution \reef{oldb3} will also appear in the
calculation of the holographic EE when the boundary theory is deformed
by a relevant operator. However, in the following, we will focus on new
contributions related to the relevant deformation.

To begin, we introduce polar coordinates $\sum_i
(dx^i)^2=dr^2+r^2\,d\Omega^2_{d-2}$ for the boundary directions in the
bulk metric (\ref{back2}). We describe the bulk surface $\m$ with a
profile $r=r(z)$ with the boundary condition $r(z=0) =R$. Then induced
metric on $\m$ is given by
 \be
 h_{\al\bt}\,dx^{\al}dx^{\bt}=\frac{L^2}{z^{2}}\left[(f(z)\, r'^2+1)dz^2+f(z)\,
 r^2\,d\Omega^2_{d-2} \right]
 \,,
  \labell{eqn:ind-metric-disk}
 \ee
where the `prime' denotes a derivative with respect to $z$. The desired
profile is chosen to minimize the area
 \bea
 A(\m)=\int dz\, d\Omega_{d-2} \sqrt{\mathrm{det}h_{\al\bt}}
 =L^{d-1}\Omega_{d-2}\int_\delta dz {f^{d/2-1}r^{d-2} \over z^{\, d-1}}\sqrt{f  r'^2+1}
 \labell{EEdisk}
 \, ,
 \eea
where $\Omega_{d-2}$ denotes the area of a $(d-2)$-dimensional unit
sphere, \ie $\Omega_{d-2}={2 \pi^{d-1 \over 2}/\Gamma\! \big({d-1 \over
2}\big)}$. As before, we introduce a UV regulator surface at
$z=\delta$.

In a pure AdS background, \ie $f(z)=1$, the profile which extremizes
the area \reef{EEdisk} has a simple form \cite{rt1,rt2}
 \be
r(z)=\sqrt{R^2-z^2}\equiv r_0\,.
 \labell{soluble}
 \ee
Unfortunately, we could not find a closed form solution in the
background with a generic relevant deformation. Hence to extract the
universal contribution, we can proceed by solving the corresponding
Euler-Lagrange equation order by order in $z$ and then substitute the
results back into the area functional \reef{EEdisk}. However, to
leading order in this expansion $f(z)=1$, for which $r=r_0(z)$ is an
exact solution. Hence it will be convenient to organize our
calculations by expanding around this profile, \ie to evaluate
corrections, $\delta r = r - r_0$, induced by the higher order terms in
$f(z)$.

In the discussion towards the end of section \ref{rest}, we found that
the small-$\tau$ expansion of the extremal area produced a series
involving powers $\tau^{n+m\frac\al2}$. In the present notation then,
we expect the small-$z$ expansion to produce terms with powers
$z^{2n+m\al}$. Further, from eq.~\reef{EEdisk}, we see that the leading
term in the integrand begins with $1/z^{d-1}$ and so to produce a
logarithmic contribution the expansion must contain a term where $2n +
m\al=d-2$. In fact, for even $d$, one finds a term where $m=0$ and
$n=(d-2)/2$ which yields the same universal contribution which appears
without the relevant deformation, as shown in eq.~\reef{oldb3}.
However, we are interested in the new contributions related to the
deformation and so where $m$ is nonvanishing --- as usual, this
requires $m\ge2$. Hence let us consider some explicit examples for
specific values of $m$.

\subsubsection{$m=2$}
\label{subsec:disk1}

In this case, we only keep the first correction $k=2$ in the expansion
of $f$, given in eq.~(\ref{series}). Note then that the cubic, as well
as any higher order interactions in the potential of the bulk scalar
play no role. Expanding eq.~(\ref{EEdisk}) to linear order in $\delta
f=f-1$ and $\delta r = r- r_0$ yields
 \be
A(\m)=L^{d-1}\Omega_{d-2}\int_\delta dz {r_0^{d-2} \over z^{\,
d-1}}\sqrt{ r_0'^2+1}\( 1+ {d-2+(d-1) r_0'^2 \over 2( r_0'^2+1)}\delta
f +\ldots \)
 \labell{ghost2}
 \ee
Above the term linear in $\delta r$ vanishes, as it must since it is
proportional to the equations of motion for $r_0$. Focusing our
attention on the $\delta f$ term above, we substitute the leading term
from eq.~(\ref{series}), as well as a small-$z$ expansion of $r_0$,
which combine to yield
 \bea
 A(\m)&\simeq&-{d-2 \over 8(d-1)} \Omega_{d-2} L^{d-1} R^{d-2}
 \int_\delta \frac{dz}{z^{d-1}}\ \left(\phi^{(0)}\frac{z^\al}{L^\al}\right)^2
 \labell{EEdisk1}\\
&&\quad
 \times\ \Big(1-{(d-4)(d-1)\over 2(d-2)}{z^2 \over R^2}
 \,+\,{(d-1)(d-3)(d-6) \over 8(d-2)}{z^4 \over
 R^4}\,+\ldots\Big)
 \nonumber
 \eea
Now if $\al=(d-2)/2$ then only the first term in the parenthesis
contributes to give a logarithmic divergence, and we obtain
 \be
S(V_\mt{S})= {\pi \over 4} {d-2 \over d-1}{L^{d-1}\over
\lp^{d-1}}\,\lambda^2\, \Omega_{d-2}\, R^{d-2}\mu^{d-2}\,\log
\mu\delta+\cdots\,,
 \labell{EEdisk2}
 \ee
where we have used eq.~\reef{convertp} to write $(\phi^{(0)})^2/L^{d-2}
=\la^2 \mu^{d-2}$. This result \reef{EEdisk2} is essentially the same
as that in eq.~(\ref{EEbelt1}). In our holographic construction, the
similarity of the results reflects the fact that to leading order
$r_0(z)\simeq R$ is constant and there is no distinction between a flat
or a spherical entangling surface. Comparing eqs.~\reef{EEbelt1} and
\reef{EEdisk2}, it appears this universal contribution can be written
in the general form:
 \be
S_\mt{univ}= {\pi \over 4} {d-2 \over d-1}{L^{d-1}\over
\lp^{d-1}}\,\lambda^2\, {\cal A}_{d-2}\,\mu^{d-2}\,\log \mu\delta\,,
 \labell{EEdiskmm}
 \ee
where ${\cal A}_{d-2}$ is the area of the entangling surface. Again
this logarithmic term will arise for any odd or even $d\ge3$, for a
relevant deformation with conformal dimension $\Delta=\frac{d}2+1$.

Given eq.~\reef{EEdisk1}, we can also begin to consider contributions
arising from terms in the expansion where $n$ is also nonvanishing. If
we consider the second term in the parenthesis in eq.~\reef{EEdisk1},
we see a new logarithmic contribution will appear if $\al=(d-4)/2$. In
this case,
 \be
S(V_\mt{S})= - \pi{ d-4  \over 8}  {L^{d-1}\over \lp^{d-1}}\,\la^2\,
\Omega_{d-2}\,\left(R\,\mu\right)^{d-4}\,\log \mu\delta+\cdots\,
 \labell{EEdisk2a}
 \ee
where we use $(\phi^{(0)})^2/L^{d-4} = \mu^{d-4}$, as implied by the
present choice of $\alpha$. Of course, this result only applies for
$d\ge5$. As should be evident from our construction above, the terms
with nonvanishing $n$ appear in the expansion of the area integrand
because of the curvature of the sphere. That is, the background
geometry has vanishing curvature, both the intrinsic and extrinsic
curvatures of the entangling surface are non-vanishing here. For
example, the Ricci scalar of the intrinsic geometry on the entangling
surface $S^{d-2}$ is given by $\R=(d-2)(d-3)/R^{2}$. This suggests that
we might express the result in eq.~\reef{EEdisk2a} as an integral over
the sphere (contributing a factor of $\Omega_{d-2}\,R^{d-2}$) but the
integrand would be $\mu^{d-4}$ multiplying some appropriate combination
of curvatures (contributing a factor of $1/R^2$).
However, given the large amount of symmetry in the present geometry, it
is not possible to precisely fix that latter curvature expression. We
continue to investigate this question in sections \ref{curve} and
\ref{pbhmatter}.

Of course, it is also possible to continue with examining higher order
terms in the expansion in eq.~(\ref{EEdisk1}). This would in turn lead
to logarithmic contributions proportional to higher powers of
curvature. Schematically, these terms would take the form
 \be
S_\mt{univ} \simeq {L^{d-1}\over \lp^{d-1}}\ \la^2\ \Omega_{d-2}
\left(R\,\mu\right)^{d-2-2n}\,\log \mu\delta\,, \labell{EEdisk2c}
 \ee
for $\alpha=(d-2-2n)/2$. Following the discussion above, it appears
that these contributions take the form of an integral over the
entangling surface with a factor of $\mu^{d-2-2n}$ multiplying some
combination of curvatures contributing a factor of $1/R^{2n}$.

\subsubsection{$m=3$} \label{firtree2x}

In this case, we are focusing on new contributions which might come
from the $k=3$ term in the expansion of $f$, given in
eq.~(\ref{series}). Again we consider the linear expansion given in
eq.~\reef{ghost2} but substitute the $k=3$ term for $\delta f$. A new
logarithmic contribution arises when we assume that $\al=(d-2)/3$. Of
course, this is the same exponent that appeared in section
\ref{firtree} and the contribution identified here has essentially the
same form as in eq.~\reef{EEbelt1a}. We combine these results to write
a general expression,
  \be
S_\mt{univ}= -{\pi\kappa \over 3(d-1)}{L^{d-1}\over \lp^{d-1}}\,\la^3\,
{\cal A}_{d-2}\,\mu^{d-2}\,\log \mu\delta\,,
 \labell{EEdiskmm2}
 \ee
where ${\cal A}_{d-2}$ is again the area of the entangling surface.
Such a logarithmic term generically apppears for any odd or even
$d\ge3$ when conformal dimension of the relevant deformation is
$\Delta=\frac{2}3(d+1)$. Comparing eqs.~\reef{EEdiskmm} and
\reef{EEdiskmm2}, we see that these two expressions have essentially
the same structure, however, the details of the overall factors differ.
In particular, the present contribution depends on the cubic coupling
in the potential for the bulk scalar. Hence it will vanish for a free
scalar field or more generally where the bulk theory is symmetric under
$\Phi\to -\Phi$.

As above, we can also consider higher order terms in the expansion with
nonvanishing $n$ which introduce additional factors of the curvature of
the sphere (\ie factors of $1/R^{2n}$) in the logarithmic
contributions. Schematically these terms again take a form very similar
to that found in the previous analysis. In particular, for
$\alpha=(d-2-2n)/3$, there are universal contributions of the form
 \be
S_\mt{univ} \simeq \kappa\,{L^{d-1}\over \lp^{d-1}}\ \la^3\
\Omega_{d-2}\! \left(R\, \mu\right)^{d-2-2n}\,\log \mu\delta\,,
\labell{EEdisk2cc}
 \ee
similar to those in eq.~\reef{EEdisk2c}. We might also note that in
particular cases the universal term receives contributions from more
than one of the expressions outlined above. For example, consider the
special case where $\al=2$ and $d=8$
--- note that this corresponds to $\Delta=6$, which is the dimension of
a scalar mass term in eight dimensions. With this choice of parameters,
we satisfy both $\al=(d-2)/3$ and $\al=(d-4)/2$ as required for the
appearance of the contributions in eqs.~\reef{EEdiskmm2} and
\reef{EEdisk2a}, respectively. Hence the full logarithmic term in the
holographic EE combines both of these contributions with
 \be
S(V_\mt{S})= - \pi {L^{7}\over \lp^7} \,  \Omega_{6}\,R^{6} \left[
{\kappa \,\lambda^3\over 21}\,\mu^6 +{ \lambda^2 \over 2 }{\mu^4\over
R^2}\]\log \mu\delta
 +\cdots\,.
 \ee

\subsubsection{$m=4$}

In this case, we extend the expansion \reef{series} of $f$ to the order
$k=4$ and hence for consistency, we must also expand the area
(\ref{EEdisk}) to quadratic order in $\delta f=f-1$. In fact, we extend
the latter expansion to quadratic order in both $\delta f$ and $\delta
r = r- r_0$ to produce
 \begin{align}
A(\m)=&L^{d-1}\Omega_{d-2}\int_a dz   {r_0^{d-2} \over z^{\, d-1}}
\sqrt{  r_0'^2+1}\Big( 1+ {d-2+(d-1) r_0'^2 \over 2( r_0'^2+1)}\delta f +g_0(z)\delta f^2
  \labell{cv}\\
  &\ \ +g_1(z)\delta f\delta r + g_2(z)\delta r^2 + g_3(z)(\delta r')^2 + g_4(z)
  \delta f \delta r' +g_5(z) \delta r \delta r' +\cdots \Big) \, ,\nonumber
 \end{align}
with
 \bea
 g_0(z)&=&{(d-2)(d-4)+2(d-2)(d-3) r_0'^2+(d-1)(d-3) r_0'^4 \over 8( r_0'^2+1)^2} \, ,
 \non
 g_1(z)&=&{(d-1)(d-2) r_0'^2+(d-2)^2 \over 2r_0( r_0'^2+1)} \, ,
 \non
 g_2(z)&=&{(d-2)(d-3) \over 2 r_0^2}
 \, , \quad
 g_3(z)={1 \over 2( r_0'^2+1)^2} \, ,
 \non
 g_4(z)&=&{(d-1) r_0'^3+d r_0' \over 2( r_0'^2+1)^2}
 \, , \quad
 g_5(z)={(d-2)r_0' \over r_0( r_0'^2+1)} \, .
 \eea
Next we must solve to the extremal profile by varying the above
`action' with respect to $\delta r$. The solution of the resulting
equation of motion must also satisfy the boundary condition $\delta
r(z=0)=0$. To leading order in $z$, we find
 \be
\delta r = {d(\al-1)+2 \over 8(d-1)(1+\al)(d-2-2\al)}
\left(\phi^{(0)}\frac{z^{\al}}{L^\al}\right)^2 {z^2\over R}+\cdots\,.
 \ee
Hence we have $\delta r\sim z^2\delta f$. Therefore if we choose
$\al=(d-2)/4$ only terms proportional to $\delta f,\delta f^2$ in
eq.~\reef{cv} contribute to the logarithmic divergence. Assuming
further that neither $(d-2)/2$ nor $3(d-2)/4$ is an integer,
eq.~\reef{cv} yields essentially the same result as in section
\ref{firtree2} and we combine them into a general expression
 \be
 S_\mt{univ}=\[{ (3d+34) \over (d-2)(d+6)^2 }\kappa^2-{(3d+8)(d-2)^2
 \over 512 (d-1)}  \] {\pi \over (d-1)} {L^{d-1} \over
 \lp^{d-1}}\la^4\mathcal{A}_{d-2}\mu^{d-2}\log \mu \delta \, ,
 \labell{EEdisk3}
 \ee
where as before $\mathcal{A}_{d-2}$ is the area of the entangling
surface. Such a logarithmic term appears for any odd or even $d\geq 3$
when $\Delta=(3d+2)/4$.

If either $(d-2)/2$ or $3(d-2)/4$ is an integer, then there are extra
terms arising from the expansion of the coefficient in front of $\delta
f $  in eq.\,\reef{cv}. These terms are associated with the effect of
intrinsic curvature of the sphere and in general will be of the form
given by eqs.~\reef{EEdisk2c} and \reef{EEdisk2cc}. Furthermore, by
suitably changing the value of $\al$, one can also consider possible
scenarios where terms involving $\delta r$ start contributing to the
universal divergence.

As a specific example, let us consider the case of a ten-dimensional
CFT with $\al=2$, which corresponds to the deformation of the CFT with
a scalar mass term. Then the logarithmic divergence is given by the
expression in eq.~(\ref{EEdisk3}) supplemented with
 \be
\delta S(V_\mt{S})={\pi \over 4} {L^{9}\over \lp^{\,9}}R^{8}\Omega_{8}
\left[ {7\,\la^2 \over 2}\, {\mu^4 \over R^4}+{\kappa\la^3 \over 3}\,
 {\mu^6 \over R^2} \right]\log \mu\delta\,+\ldots
 \ee
We have used eqs.~\reef{series}, \reef{coeffs} and \reef{ghost2} to
evaluate this term.

\section{Curved boundaries} \label{curve}

In section \ref{sec:disk}, we examined the holographic EE for a
spherical entangling surface. Our calculations there began to
illustrate an interesting interplay in the coefficient of the universal
contributions between the curvature of the entangling surface and the
mass scale introduced by the relevant deformation. In particular, our
results suggest that various new universal contributions to the
holographic EE appear where the coefficient is given by an integral
over the entangling surface with a factor of $\mu^{d-2-2n}$ multiplying
some combination of curvatures contributing a factor of $1/R^{2n}$.
However, with the results of the previous section alone, the details of
these contributions remain incomplete. That is, the precise form of the
appropriate curvature factor remains unclear. Here we examine these
issues further by calculating the holographic EE for various entangling
surfaces when the background in which the boundary theory resides is
also curved.

In particular, we consider the boundary theory on certain simple
backgrounds of the form $R^1 \times \Sigma_{k}$ where $\Sigma_k$ is a
maximally symmetric space, where $k \in \{\pm 1,0\}$ indicates the sign
of the curvature. That is, $\Sigma_{+}=S^{d-1}$, $\Sigma_0=R^{d-1}$ and
$\Sigma_-=H^{d-1}$. Further, we introduce $R$ as the background
curvature scale so that the Ricci scalar takes the form
 \be
R[\Sigma_{k}] = \frac{k(d-1)(d-2)}{R^2}\,.\labell{Rscalar}
 \ee
Of course, $R$ can be scaled away in the case of $k=0$ --- the simplest
choice is to set $R=L$ in this case. The corresponding bulk metric can
be written as
 \be
ds^2 = \frac{L^2}{z^2}\left(dz^2 - f_t(z) dt^2 + R^2\, f_k(z)\,
d\Sigma_{k}^2 \right)\,, \labell{bulkmx}
 \ee
where
 \be
d\Sigma_{k}^2 = d\theta^2 + F_k(\theta)^2 d\Omega_{d-2}^2\,, \qquad F_k
= \left\{ \begin{array}{cc}
\sin\theta\,, & \ k=1\\
\sinh\theta\,, & \ \ \ \,k=-1\\
\theta \,, & \ k=0
\end{array}\right.\labell{hoot}
 \ee
and $d\Omega_{d-2}^2$ is the metric on a ($d-2$)-dimensional unit
sphere. In a pure AdS background, the two metric functions $f_{t,k}(z)$
are given by:
 \be
 f_t= \left(1 + k \frac{z^2}{4 R^2}\right)^2\equiv f_{t,0}\,, \qquad
 f_k = \left(1 - k \frac{z^2}{4 R^2}\right)^2 \equiv f_{k,0}\,.
 \labell{pureA}
 \ee

We wish to consider the Einstein-scalar theory \reef{action} with the
cubic potential \reef{pot2}, as in the previous section. The Einstein
equations \reef{Einstein} now yield three nontrivial components but
only two of these are independent. We chose to consider the following
two equations:
 \bea
\frac{( d-2) (d-1)}{2}\bigg[\left(\frac{f_k'}{f_k}\right)^2
-\frac{4}{z} \left(\frac{f_k'}{f_k}\right) \bigg]
&&+(d-2)\frac{f_k'f_t'}{f_k f_t} - 2(d-1)\frac{f_t'}{z f_t} \nonumber \\
&&- 2 \frac{R[\Sigma_{k}]}{f_k}
-\Phi'^2 + \frac{(mL)^2}{z^2}\Phi^2 + \frac{\kappa}{3z^2}\Phi^3 =0 \nonumber \\
2(d-1)\bigg[f_k''-\frac{d-1}{z}f_k'
+\frac{(d-4)}{4}\frac{f_k'^2}{f_k}\bigg]
&&- 2\, R[\Sigma_{k}] \labell{eeom2} \\
&&+ f_k\left(\Phi'^2 + \frac{(mL)^2}{z^2}\Phi^2 +
\frac{\kappa}{3z^2}\Phi^3 \right)=0 \nonumber
 \eea
where $R[\Sigma_{k}]$ is the Ricci scalar \reef{Rscalar} of the
boundary geometry. The scalar field equation \reef{scalar} becomes
 \be \Phi'' -\frac{d-1}{z}\Phi' +
\frac{\Phi'}{2}\left((d-1)\frac{f_k'}{f_k}+ \frac{f_t'}{f_t}\right)
-\frac{(mL)^2}{z^2}\Phi - \frac{\kappa}{2z^2}\Phi^2=0\,.
 \ee

In order for the bulk metric \reef{bulkmx} to be an asymptotically AdS
solution, $f_k$ and $f_t$ must approach a constant (\ie 1) at the
boundary $z\to 0$. Substituting the asymptotic form $\Phi \sim
z^\alpha$ into the scalar equation again yields the expected indicial
equation, $\alpha(\alpha-d)= (mL)^2$, which has the two solutions
$\Delta_\pm$ given in eq.~\reef{dimension}. As before, we introduce a
profile for $\Phi$ beginning with $z^{\alpha}$, where $\alpha =
\Delta_-=d/2 - \sqrt{d^2/4 + (mL)^2}$, to describe a dual operator of
dimension $\Delta_+ = d-\alpha$. The back-reaction on the metric, \ie
in $f_{k,t}$, again begins at order $z^{2\alpha}$. However, the
boundary curvature now also appears as an explicit source in the
Einstein equations (\ref{eeom2}) and its effect begins to appear at
order $z^2$. The asymptotic expansion of the metric thus generally take
the form given in eq.~\reef{expandgx}.

Now we would like to compute the EE in the case where the entangling
surface is an $S^{d-2}$ at $\theta=\theta_0$ in the metric \reef{hoot}
for the spatial geometry $\Sigma_{k}$. Hence we specify the bulk
surface with a profile $\theta(z)$ satisfying the boundary condition
$\theta(z=0)=\theta_0$. This calculation then requires a generalization
of eq.~\reef{EEdisk},
 \be
S = \frac{2\pi}{\lp^{d-1}} L^{d-1} R^{d-2}\,\Omega_{d-2}\,\int_\delta
dz \frac{f_k^{d/2-1}F_k^{d-2}}{z^{d-1}} \sqrt{1+ f_k\,R^2
\theta'(z)^2}\,. \labell{EEC}
 \ee
In a pure AdS background \reef{pureA}, we can find an exact solution
for $\theta(z)$:
 \be
\theta(z)\equiv\theta_{k,0}(z) = \left\{ \begin{array}{cc}
\cos^{-1}\left( \cos\theta_0 (4R^2+z^2)/(4R^2-z^2)\right)\,, & \ \ k=1\\
\cosh^{-1}\left( \cosh\theta_0 (4R^2-z^2)/(4R^2+z^2)\right)\,, &
 \ \ \ \ k=-1\\
\sqrt{\theta_0^2 - z^2/R^2}\,, & \ \ k=0
\end{array}\right.\labell{exactt}
 \ee
We were able to find these solutions because these profiles all specify
essentially the same surface in different coordinate systems of the AdS
geometry. Following ref.~\cite{circle4}, this surface corresponds to
the bifurcation surface of a topological AdS black hole.

Now we follow the same procedure as in section \ref{sec:disk} expanding
around the pure AdS solutions. That is, we expand eq.~\reef{EEC} in
powers of $\delta f$ and $\delta \theta$, which are defined as
 \be
\delta f = f_k- f_{k,0}\,,\qquad \delta \theta = \theta-
\theta_{k,0}\,.
 \ee
Note that $f_t(z)$ does not appear in our integral \reef{EEC} and so we
need not consider perturbations of this metric function. To obtain the
leading contribution in $\phi^{(0)}$, we expand eq.~\reef{EEC} to
leading order in $\delta f$ and $\delta \theta$, which gives
 \bea
 \delta S\simeq 2\pi\frac{ L^{d-1}}{\lp^{d-1}}  R^{d-2}\Omega_{d-2}&&\int_\delta
 dz \left[\frac{f_{k,0}^{d/2-1}\,F_k^{d-2}(\theta_{k,0})}{z^{d-1}}
\sqrt{1+ R^2\,f_{k,0}\, \theta_{k,0}'{}^2}\right.
\nonumber \\
&&\times\left.\left(\frac{d-2}{2 f_{k,0}} + \frac{\theta_{k,0}'{}^2}{1+
R^2\,f_{k,0}\,\theta_{k,0}'}\right)\right]\,\delta f
\nonumber \\
\simeq \pi(d-2)\frac{L^{d-1}}{\lp^{d-1}}&&R^{d-2}
\Omega_{d-2}\,F_k(\theta_0)^{d-2} \int_{\delta} \frac{dz}{z^{d-1}}
 \,\delta f   \nonumber \\
&&\times \left[1  -\frac12(d-4)\left(\frac{d-1}{d-2}\,  c_k^2 +
\frac{k}2 \right) \frac{z^2}{R^2}\right] \labell{owl2}
 \eea
where
 \be c_k = \left\{
\begin{array}{cc}
\cot\theta_0\,, &\ \  k=1\\
\coth\theta_0\,, &\ \ \ \  k=-1\\
\theta_0^{-1} \,, &\ \  \,k=0\,.
\end{array} \right.
 \ee
Note that the term linear in $\delta \theta$ vanishes in
eq.~\reef{owl2} by the equations of motion (for the extremal profile in
AdS space).

Following the discussion in section \ref{notes}, the correction to the
metric $\delta f$ may be written as
 \be \delta f =
\sum_{n=0}^{N-1}\sum_{m=2}^\infty a_{(m,n)}
\left(\phi^{(0)}(z/L)^\alpha\right)^{m} (z/R)^{2n}\,,
 \labell{expandfx}
 \ee
where we have assumed the exponent $\alpha$ has the form
$\alpha/2=N/M$, as in the previous analysis. Our convention to
normalize the factors of $z^{2n}$ with powers of $R$, rather than $L$,
is convenient in the following but it is also a natural choice
because $R[\Sigma_{k}]$ appears as a source in the
Einstein equations \ref{eeom2}. The leading coefficient $a_{(2,0)}$ in
this expansion is unaffected by the boundary curvature and takes
precisely the same value as in eq.~\reef{coeffs}, \ie
$a_{(2,0)}=a_2=-1/[4(d-1)]$.

Consider first the universal contribution from $\delta S$ arising when
$\alpha =(d-2)/2$, as appeared in sections \ref{eme2a} and
\ref{subsec:disk1}. In this case, the leading contribution in $\delta
f$ is of order $z^{d-2}$ and the new logarithmic term is identical to
that in eq.~\reef{EEdiskmm}. In particular then, this result is
unaffected by the background curvature.

Next we turn to $\alpha=(d-4)/2$, as was considered in sections
\ref{firtree} and \ref{firtree2x}. In this case, $\delta f$ begins at
$z^{d-4}$ but must be expanded up to $z^{d-2}$ to identify the
logarithmic contribution to $\delta S$. The equations of motion give
 \be
\delta f = (\phi^{(0)})^{\,2}
\left(\frac{z}{L}\right)^{2\alpha}\left[-\frac{1}{4(d-1)} + k\,\frac{(
d-4) (d^2 - 4 d + 8) }{32 (d-2)(d-1) } \frac{z^2}{R^2} \right] +
\cdots\,.
 \ee
The logarithmic contribution in eq.~\reef{owl2} then becomes
 \bea
S_\mt{univ}&=&\pi (d-2)\frac{ L^{d-1}}{\lp^{d-1}}\,R^{d-4}\,
\Omega_{d-2}\,F_k(\theta_0)^{d-2}\,\lambda^2\,\mu^{d-4}\,
\log\mu\delta \nonumber \\
&& \qquad\times\  \left[ \frac12(d-4)\left(\frac{d-1}{d-2}\,  c_k^2 +
\frac{k}2 \right)\, a_{(2,0)}-a_{(2,2)}
\right] \nonumber \\
&=& -\pi\frac{L^{d-1}}{\lp^{d-1}}\,
\lambda^2\,\mu^{d-4}\,
\log\mu\delta\ \int_{S^{d-2}}d^{d-2}\sigma\,\sqrt{H} \labell{Scurve} \\
&&\qquad\times\ \left[\frac{(d-4)}{8(d-2)^2}\,(K^{\hat\theta\,a}_{a})^2
+ \frac{(d-4)(d^2-2d+4) }{32(d-1)^2(d-2)
}\,R[\Sigma_{k}]\right]\,.\nonumber
 \eea
In the second expression, we have tentatively expressed the result as
an integral over the entangling surface, to illustrate the kind of
general expression that we anticipate. We are denoting the induced
metric on this boundary surface as $H_{ab}$. Note that implicitly the
result contains two curvature scales, $1/R^2$ and $c_k^2/R^2$ and hence
the integrand includes two independent curvature terms. The last term
involves the Ricci scalar \reef{Rscalar} of the background geometry in
which the boundary theory resides. The first term involves the
extrinsic curvature of the entangling surface which is given by
 \be
K^{\hat\theta}_{ab}= - t^i_a t^j_b\, \nabla_i\, n^{\hat{\theta}}_{j} =-
\frac{c_k}{R}\,\hz_{ab}\,,
 \labell{extrinsic}
 \ee
where $n^{\hat{\imath}}_j$ and $t^i_a$ are respectively the normal and
tangent vectors, to the entangling surface $\Sig$ --- see \cite{EEGB}
for further details and a full discussion of our conventions. Note that
in principle, there is also an extrinsic curvature associated with the
normal vector in the time direction however $K^{\hat t}_{ab}=0$ in the
present case.

Hence our present calculation demonstrates that $S_\mt{univ}$ takes a
form slightly more complicated than anticipated in the discussion in
section \ref{subsec:disk1}. In particular, there are two independent
curvature contributions, whereas we could only detect one in our
calculations in the previous section. We should note however that the
curvatures which we have written in eq.~\reef{Scurve} are only
representative. For example, we easily could replace
$(K^{\hat\theta\,a}_{a})^2$ by $(d-2)\,
K^{\hat\theta\,a}_{b}\,K^{\hat\theta\,b}_{a}$. Alternatively we could
use the fact that the intrinsic curvature of our entangling surface has
$\R\propto(1/R^2\ +\ c_k^2/R^2)$. Of course, when we set $k=0$ in
eq.~\reef{Scurve}, the result agrees with this previous calculation in
a flat background. While they are informative, unfortunately these
simple examples are still too symmetric to give us enough insight to
properly fix the covariant expression that describes this universal
term for a general entangling surface.

\section{PBH transformations with matter} \label{pbhmatter}

In this section, we revisit the powerful approach developed in
\cite{adam} to get a more precise understanding of the new universal
contributions to the holographic EE. Here one is able to determine
essentially all of the fixed boundary data by examining their behaviour
under PBH transformations, the subgroup of bulk diffeomorphisms which
generate Weyl transformations in the boundary. In \cite{adam} however
only pure gravity theories in the bulk are considered and so we must
extend their analysis to include a bulk scalar. An essential feature of
our analysis is that we must not just consider $\phi^{(0)}$ to be a
coupling constant in the boundary theory, rather we must elevate it to
a field. That is, we consider $\phi^{(0)}(x)$ to take full advantage of
this approach. Just as in the pure gravity case, these calculations
leave some undetermined constants that must be fixed by the equations
of motion. While there are no immediate obstacles to performing a
general analysis, in the following, we only work out a specific example
which includes a scalar field in the bulk to illustrate the general
approach.

In particular, we will focus our attention on $\alpha =(d-4)/2$ and
completely fix the universal contribution which was identified in the
previous section. We fix the metric in FG gauge as in eq.~\reef{back0},
and define
 \be
\tilde{g}_{ij}(\phi^{(0)})= g_{ij} + \Delta g_{ij} (\phi^{(0)})\,,
 \ee
where $g_{ij}$ is the asymptotically AdS solution without the relevant
deformation turned on, \ie before the back-reaction of the scalar field
is considered. Our goal is to solve for the leading terms in the
expansion of $\phi_{(0)}$ in $\Delta g_{ij}$. The leading terms in the
expansion of the metric and the scalar field $\Phi$ are
 \bea
g_{ij}(x,\rho)&=& \gz_{ij} + \rho\go_{ij}  + \cdots\,, \qquad \Delta
g_{ij}= \rho^{\alpha} \left(\gf{\alpha}_{ij} + \rho \gf{\alpha
+1}_{ij}\right )+ \cdots
\,, \nonumber \\
\Phi(x,\rho) &=& \rho^{\frac{\alpha}{2}} \left(\phi^{(0)}+
\rho\phi^{(1)}\right)+\cdots\,.
 \eea
Now the coordinate transformations which preserve the FG gauge take the
form \cite{adam},
 \be
\rho = \rho'(1- 2\sigma(x')), \qquad x^i = x'^{i} + a^i(x',\rho')\,,
 \labell{pbh}
 \ee
to leading order in some function $\sigma$, where
 \be
a^i(x,\rho) = \frac{L^2}{2}\int_0^\rho d\rho'
g^{ij}(x,\rho')\,\partial_j \sigma(x)\,.
 \labell{diffeo}
 \ee
The form of $a^i$ is independent of the form of the series expansion of
$g_{ij}(x,\rho)$ in $\rho$, which is modified from that in
eq.~\reef{expand} to the more general form in eq.~\reef{expandgx} in
the presence of matter back-reaction. Now following the approach of
\cite{adam}, we substitute the metric expansion \reef{expandgx} into
eq.~\reef{diffeo} and use
 \be
\delta G_{i j} = \frac{\delta g_{i j}(x, \rho)}{\rho}
=\xi^\mu\,\partial_\mu G_{i j}
 + 2\partial_{(i}\xi^\mu G_{j)\mu}\,,\labell{transformg}
 \ee
where
 \be
\xi^\rho = -2\sigma(x) \rho \,, \qquad \xi^i = a^i(x,\rho)\,.
 \ee
Our notation is such that the `symmetrization bracket' is defined as
$A_{(i}B_{j)}= 1/2(A_iB_j + B_i A_j)$. This allows one to determine how
each coefficient in the general expansion \reef{expandgx} transforms
under a general PBH transformation.

One can tell immediately from $\xi^\rho \partial_\rho G_{ij}$ that
there is a homogeneous scaling term for each coefficient of the form
 \be
\delta \gf{n} = -2\sigma(x)(n-1)\gf{n} + \cdots\,.
 \ee
Since the PBH transformations reduce to Weyl rescalings in the
boundary, the above implies that
$\overset{\scriptscriptstyle{(n)}}{g}_{ij}$ has conformal dimension
$2(n-1)$. In other words, the conformal dimension can be read off from
the power of $\rho$ multiplying the coefficient of interest.
Particularly, $\overset{\scriptscriptstyle{(0)}}{g}_{ij}$ always
carries conformal dimension $-2$, as expected.

We are interested in how
$\overset{\scriptscriptstyle{(\alpha)}}{g}_{ij}$ and
$\overset{\scriptscriptstyle{(\alpha+1)}}{g}_{ij}$ transform. Expanding
$a^i(x,\rho)$ in $\rho$, \be a^i(x,\rho) = a^i_{(1)} \rho +
a^i_{(2)}\rho^2 + a^i_{(\alpha+1)}\rho^{\alpha+1} + \cdots, \ee and
substituting into eq.~\reef{transformg}, we have
 \bea
\delta \gz_{ij} &&= 2 \sigma \gz_{ij}\,, \qquad  \delta \gf{1}_{ij} =
a^k_{(1)} \partial_k
 \gz_{ij}+ 2 \partial_{(i}a^k_{(1)}\gz_{j)k} \,, \nonumber \\
\delta \gf{\alpha}_{ij}&&=-2\sigma (\alpha-1)\gf{\alpha}_{ij}\,, \labell{hose} \\
\delta \gf{\alpha+1}_{ij} &&= -2\sigma \alpha \gf{\alpha+1}_{ij}+
a^k_{(1)}\partial_k\gf{\alpha}_{ij}+ a^k_{(\alpha+1)}\partial_k\gz_{ij}+ 2
\partial_{(i}a^k_{(1)}\gf{\alpha}_{j)l} + 2 \partial_{(i}a^k_{(\alpha+1)}\gz_{j)k}\,,
\nonumber
 \eea
where
 \be a^i_{(1)}= \frac{L^2}{2}(\gz^{-1})^{ij}\,\partial_j\sigma \,, \qquad
a^i_{(\alpha+1)}= -\frac{L^2}{4}(
\gz^{-1}\gf{\alpha}\,\gz^{-1})^{ij}\,\partial_j\sigma\,.
 \ee
These give the Weyl transformation properties of the coefficients, with
which one could in principle reconstruct the series. The building
blocks in the boundary theory considered in \cite{adam} include the
boundary metric, its curvature tensors and their covariant derivatives.
The only extra component that we have at our disposal here is the
nontrivial boundary source $\phi^{(0)}(x)$ of conformal dimension
$\alpha=d-\Delta$ and its covariant derivatives.

The solution for $\overset{\scriptscriptstyle{(1)}}{g}_{ij}$ is
unaffected by the scalar profile and is again given by
eq.~\reef{metricexpand}. Meanwhile
$\overset{\scriptscriptstyle{(\alpha)}}{g}_{ij}$ transforms
homogenously with conformal dimension $2\alpha -2$. Including
$\phi^{(0)}$ amongst our building blocks, the solution is uniquely
determined as
 \be
\gf{\alpha}_{ij} =  c_1\,(\phi^{(0)})^2\, \gz_{ij}\,, \labell{g1}
 \ee
where the constant $c_1$ is fixed by the bulk equations of motion.

Substituting $\overset{\scriptscriptstyle{(\alpha)}}{g}$ back into
$a^i_{(\alpha+1)}$ and hence the transformation of
$\overset{\scriptscriptstyle{(\alpha+1)}}{g}$, we find the latter must
have the form
 \bea
\gf{\alpha+1}_{ij}& = &c_1\,L^2\,\left(d_1 (\phi^{(0)})^2 R_{ij}+ d_2
\gz_{ij}(\phi^{(0)})^2R + d_3\,\partial_i \phi^{(0)}\partial_j
\phi^{(0)}\right. \labell{g11}\\
&&\qquad\qquad\qquad\left.+\ d_4\nabla_{i}\nabla_j(\phi^{(0)})^2 +d_5
\gz_{ij}\Box (\phi^{(0)})^2\right)\,.\nonumber
 \eea
Here we have more degrees of freedom than equations, and we obtain
 \bea
d_1&=& -\frac{(d-4)(d+ 2d_5(d^2-8d+12))}{2(d-2)^2}\,, \qquad
d_2=\frac{(d-8d_5(d-2))(d-4)}{4(d-2)^2(d-1)}\,, \nonumber \\
d_3&=&-\frac{2(d^2-5d+4 + 2d_5(d^3-11d^2+ 36 d-36))}{(d-4)(d-2)} \,,
\quad d_4=\frac{1}{2}-d_5(d-6)\,,
 \eea
leaving $d_5$ to be determined by equations of motion.

The transformation of the scalar field gives
 \bea
\delta \phi^{(0)} &&= -\sigma \alpha \phi^{(0)} \,,\nonumber \\
\delta \phi^{(1)} &&= -\sigma (\alpha+2) \phi^{(1)}+
\frac{L^2}{2}(\gz^{-1}){}^{ij}\,\partial_i\sigma\,\partial_j\phi^{(0)}\,,
 \eea
implying that
 \be \phi^{(1)}=
\frac{L^2}{2(d-2(\alpha+1))}\,
\left(\Box\phi^{(0)}-\frac{1}{2(d-1)}\phi^{(0)} R\right)\,.
 \ee
In general if  $n\alpha=2$ for some integer $n$ then an extra
homogenous term $(\phi^{(0)})^{\,n+1}$ could appear in $\phi^{(1)}$ and the
coefficient of this term would have to be determined from the
equations of motion.

As a check, our results above were compared with those obtained from
directly solving the equations of motion with $d=6,\alpha=(d-4)/2=1$ and they are
completely consistent. Note also that in addition to coefficients that
are exactly determined above, the coefficient $d_5$ is over-determined
by the equations of motion but may be consistently solved. Hence this
serves as a non-trivial check.

With these results, we can compute the leading $\phi^{(0)}$
contribution to the universal logarithmic term in the entanglement
entropy for arbitrary boundary entangling surface. The procedure is
similar to the previous section. We begin by assuming that the bulk
surface in the absence of relevant perturbation is given by
$X^\mu(x^\alpha,\tau)$ --- where we are again working with the gauge
\reef{eqn:gauge-brane}. The back-reaction of the scalar field then
introduces changes in the background metric $\Delta g$ and also in the
minimal surface $\delta X$. The former has been solved to leading order
in $\phi^{(0)}$ above. The latter however, does not contribute to the
entanglement entropy to leading order because $X^i(x^\alpha,\tau=\rho)$
extremizes the action at $\phi^{(0)}=0$, a fact we have made used of
already in the previous sections. Since $\Delta g$ begins at
$\tau^{\alpha}= \tau^{(d-4)/2}$, together with the measure of the
minimal surface $\sqrt{h}\sim \tau^{-d/2}$, the leading correction to
the entanglement entropy goes like $\tau^{-2}$. To extract the
log-term, one needs to expand the remaining integrand to linear order
in $\tau$. While we do not know the complete solution of $X(x,\tau)$
for arbitrary asymptotically AdS background and boundary entangling
surface, the linear $\tau$ term in its asymptotic expansion is
universal, completely dictated by fixed boundary data, independent of
the gravity theory concerned. That is, it can also be fixed by the PBH
transformations which yield \cite{adam}
 \be X^i(x^\alpha,\tau) =
\XX{0}^{\,\,\,i}(x^\alpha) +
 \tau \XX{1}^{\,\,\,i}(x^\alpha ) + \cdots\,,\qquad \XX{1}^{\,\,\,i}
 = \frac{L^2K^i}{2(d-2)}\,,
 \ee
where $K^i$ is the trace of the extrinsic curvature of the entangling
surface --- see eq.~\reef{extrinsic}.

The leading $\phi^{(0)}$ dependence of the area of the minimal surface
is then given by
 \bea
\delta A &&= \int d^{d-2}y d\tau\ \delta\!
\left(\sqrt{h_{\tau\tau}(\phi^{(0)})\det{h_{ij}(\phi^{(0)}}})\right)
 \label{green} \\
&&=  L\int d^{d-2}y \,\frac{d\tau}{2\tau^{d/2}}
\sqrt{\tilde{h}_{\tau\tau}\det{\tilde{h}_{ab}}}\left(\frac{2\tau
\partial_\tau X^\mu\partial_\tau X^\nu}{\tilde{h}_{\tau\tau}}
+\frac{\partial_a X^\mu\partial_b
X^{\nu}\tilde{h}^{ab}}{2}\right)\bigg\vert_{\phi_0=0}\Delta
g_{\mu\nu}\,,
 \nonumber
 \eea
where we have defined $\tilde{h}_{\tau\tau}= 4\tau^2\, h_{\tau\tau},\,\,
\tilde{h}_{ab}= \tau h_{ab}$, such that these quantities begin at
$O(\tau^0)$. Further recall that the radial integral ends at the UV
regulator surface $\tau_{min}=\delta^2/L^2$. With $\phi^{(0)}=0$, the
expansion of $\tilde{h}_{\tau\tau}$ and $\det{\tilde{h}_{ij}}$ are
given by \cite{adam}
 \be
\tilde{h}_{\tau\tau} = L^2+ 4\tau
(\XX{1}^{\,\,\,i})^{2}+\cdots\,,\qquad \det{\tilde{h}_{ij}} =
\det{\hz_{ab}}(1+ \tau \hz^{\,ab}\,\ho_{ab}+ \cdots)\,,
 \ee
where
 \be
\ho_{ab} \,=\, \go_{ab}-\frac{L^2}{d-2}K^iK^j_{ab}\gz_{ij}\,,
 \ee
and $\overset{\scriptscriptstyle{(1)}}{g}_{ab}$ is as defined in
eq.~\reef{metricexpand}, but projected on to the boundary entangling
surface by contracting with tangent vectors
$\partial_a\overset{\scriptscriptstyle{(0)}}{X}{}^i$. We finally have
 \bea
S_\mt{univ}&&= -2\pi\frac{ L^{d-1}}{\lp^{d-1}}\,\lambda^2\int d^{d-2}y
\sqrt{\hz_{ab}}\,\bigg(-\frac{(d-1)(d-4)}{4(d-2)^2}\,c_1\,K^iK^j
\gz_{ij}
\nonumber \\
&& +\frac{(d-4)}{4L^2}\,c_1\,\go_{a}^{\,\,\,\,a} +
\frac{1}{2L^2\phi^{(0)}{}^2
}\gf{\alpha+1}_{a}^{\,\,\,\,a}\bigg)\,\mu^{d-4}\,\log\mu\,\delta\,.
 \labell{full}
 \eea
This expression is now completely fixed when we apply
 \be
c_1 = -\frac{1}{4(d-1)}\,, \qquad d_5= \frac{1}{8}\,,
 \ee
which were determined by solving the equations of motion.

We can combine the preceding results to explicitly write out this
universal contribution for the case that $\phi_{(0)}$ is a constant.
With this simplification, the result \reef{full} reduces to
 \bea
S_\mt{univ}&=& -\frac{(d-4)\pi}{32(d-2)^2}\frac{
L^{d-1}}{\lp^{d-1}}\,\lambda^2\,\mu^{d-4}\,\log\mu\delta \labell{guub}\\
&&\quad\times\ \int d^{d-2}y \sqrt{\hz_{ab}}\bigg(4\,K^iK^j \gz_{ij}
 +\frac{d^2+4}{d-1}\,R_{a}^{\,\,a} -2d\frac{d-2}{(d-1)^2}\,R\bigg)\,.
 \nonumber
 \eea
where $R$ and $R_a{}^a$ are, respectively, the background Ricci scalar
and the background Ricci curvature contracted with $H_{ij}$, the
induced metric expressed as a $d$-dimensional tensor:
$H_{ij}=\overset{\scriptscriptstyle{(0)}}{g}_{ij}-n^{\hat\imath}_i\,n^{\hat\imath}_j$.
Evaluating this expression \reef{guub} for the geometries considered in
the previous section, we find complete agreement with
eq.~\reef{Scurve}. However, the present result is completely general
and can be applied for any background geometry and any (smooth)
entangling surface.

\section{Discussion} \label{discuss}

Our calculations have demonstrated two interesting properties about
holographic entanglement entropy. First of all, the coefficient of any
universal contribution which is logarithmic in the short-distance
cut-off is independent of the state of the boundary theory. Secondly,
when the boundary theory is deformed by turning on a relevant operator,
new universal contributions appear including a class of the form found
in \cite{frank}.

Let us begin here with some discussion of the first result. The
observation that these universal coefficients are independent of the
state of the underlying field theory may seem trivial. As previously
noted for an even dimensional CFT, the universal coefficients will be
given by some linear combination of central charges in a general
setting, even without holography. However, while our result is
implicitly regarded as `obvious' in discussions of EE, a rigorous proof
has not been provided. In the AdS/CFT framework, we were able to make
the separation of data depending on the state versus data depending on
the action very explicit, even when the boundary theory is deformed
by a relevant operator, and it is clear only the latter data
contributes to determining the universal terms in the holographic EE.

Let us point out that there is the potential to produce a contradiction
with relevant deformations with low conformal dimensions. Recall that
the standard approach, described in section \ref{notes}, allows us to
study $\Delta\ge d/2$. The lower bound arises with $m^2= -d^2/4L^2$,
which corresponds to the Breitenlohner-Freedman bound in $d+1$
dimensions \cite{BF}. However, the unitarity bound for a scalar
operator in a $d$-dimensional CFT allows for $\Delta\ge (d-2)/2$. To
study operators in the range $d/2\ge\Delta\ge(d-2)/2$, we must use the
`alternative quantization' of the dual bulk scalar set forward in
\cite{igor9} for masses in the regime:
 \be
-\frac{d^2}{4L^2}\le m^2\le -\frac{d^2}{4L^2}+\frac{1}{L^2}\,.
 \labell{alter}
 \ee
Hence in this regime then, we can choose the dimension of the dual
operator as $\Delta=\Delta_-$ and in this case, the roles of
$\phi^{(0)}$ and $\phi^{(\Delta-\frac{d}2)}$ are interchanged. We note,
however, that this alternate quantization does not change the powers of
$\rho$ appearing in eq.~\reef{field0}. In particular, the leading power
is still given by $\rho^{\Delta_-/2}$. However, the key difference (for
our purposes) is that the leading coefficients appearing in this
asymptotic solution are now related to the state of the boundary field
theory.

Hence it seems that in this situation any new universal term appearing
in the holographic EE must depend on the state. However, as we now
show, there is no problem because deformations in this regime do not
produce any such universal contributions. Recall from our discussion in
section \ref{rest} that a universal contribution appears in the
holographic EE when, in the expansion of the integrand in
eq.~\reef{arealx}, a term appears with
$\tau^{n+m\frac{\alpha}{2}}=\tau^{\frac{d-2}2}$. Further, recall that
apart from $m=0$, the minimum value of $m$ is 2 because of the
structure of the Einstein-scalar theory in the bulk. This means that
there is a maximum value which $\alpha$ can have in order to produce a
logarithmic contribution in the holographic EE. In particular, a
logarithm will only arise for $\alpha\le\frac{d}2-1$. In terms of the
conformal dimension of the boundary operator, this corresponds to
$\Delta\ge \frac{d}2 +1$ or in terms of the mass of the bulk scalar,
 \be
 m^2\ge -\frac{d^2}{4L^2}+\frac{1}{L^2}\,.
 \labell{limit}
 \ee
However, the lower limit here is interesting because comparing to
eq.~\reef{alter}, we see that it precisely excludes the range of
allowed masses where it is possible to make an alternate quantization
of the bulk scalar.\footnote{The case of $\Delta=(d-2)/2$ may still
seem problematic because it corresponds to precisely the limit $m^2=
-d^2/(4L^2)\,+\,1/L^2$. However, this is precisely the unitarity bound
for which the dual operator is expected to be a free scalar field.
However, such a CFT would be beyond the scope of the holographic models
which we are considering here.} Hence the potential problem, arising
from the interchange of the roles of the different terms in the
asymptotic scalar in the alternate quantization, is cleanly avoided
because the deformation will simply not generate a $\log\delta$
contribution in the holographic EE.

While the focus of our discussion has been the possible logarithmic
contributions to the holographic EE, these are only the least divergent
terms as $\delta\to0$. The expansion of the area \reef{arealx} will
generally produce a series of terms diverging as
$1/\delta^{d-2-2n-m\alpha}$. Of course, the first term (\ie with
$n=0=m$) yields the expected area law. Further, our analysis shows that
the coefficients of all of these divergent terms are determined by the
fixed boundary data in the asymptotic expansion, \ie they are all
independent of the state of the boundary theory. In the present
holographic framework, the general coefficient will contain $m$ factors
of the coupling $\phi^{(0)}$ and an integral of $n$ curvatures over the
entangling surface. This observation then guarantees that the mutual
information
 \be
I(A,B)=S(A)+S(B)-S(A\cup B) \labell{mutual}
 \ee
for two disjoint regions $A$ and $B$ is free of any UV divergences in
our holographic calculations. The finiteness of the mutual information
is another generally accepted feature which is believed to be true in
general but never rigorously proven.

We should add that implicitly we are considering a constrained class of
states in this discussion,\footnote{RCM thanks Mark van Raamsdonk for
an interesting conversation on this point.} \eg the energy density of
the states being studied must be kept finite. This constraint becomes
evident with the following thought experiment: Consider the boundary
theory with a finite cut-off $\delta$, in which case it contains a
finite number of degrees of freedom (if the total volume is also kept
fixed). In this case, one can easily imagine choosing a state in which
there is simply no entanglement between a particular region $V$ and its
complement $\bar V$. That is, we seem to have removed the potentially
divergent contributions to the entanglement entropy with a particular
choice of state. However, the price to be paid for this lack of
correlations would be that the energy density of such a state will be
of order $1/\delta^d$. Hence if we wish to maintain this vanishing
entanglement in the limit $\delta\to0$, we would require an infinite
entanglement entropy. Holographically, such a state would not be dual
to an asymptotic AdS geometry and so it lies outside of the class of
states considered here.

In the discussion above eq.~\reef{inducemet}, we noted that the
analysis in refs.~\cite{grwit,adam} provided a general analysis for
bulk submanifolds with an arbitrary dimension $k+1$. In this case, the
second set of independent coefficients appear in the expansion of the
embedding functions at order $\tau^{(k+2)/2}$. Our analysis focussed on
$k=d-2$, however, for smaller values of $k$, the second set of free
coefficients would appear at a lower order than in the expansion given
in eq.~\reef{expandx}. Despite appearing at a lower order, this state
data does not contribute to the coefficients of any UV divergences, in
particular a logarithmic divergence, appearing in the calculation of
the area of the corresponding surfaces. This occurs precisely because
the dimension of the submanifold is also reduced. Hence when we
evaluate the analog of eq.~\reef{arealx}, the leading power becomes
precisely $\tau^{-(k+2)/2}$ and so the state dependent coefficients
will only produce finite contributions to the area for general $k$.
Hence our results extend beyond the calculation of the entanglement
entropy. For example, this analysis would apply to the calculation of
the expectation values of Wilson lines and shows that the coefficients
of any divergent terms appearing in such a calculation are also
independent of the state of the boundary theory.

In certain cases, no logarithmic contribution appears in the EE, \eg
with a CFT in an odd number of spacetime dimensions. However, the
constant term independent of the short-distance cut-off may then still
be a universal contribution to the EE \cite{rt1,rt2}. The universality
of this constant contribution is established for a variety of $d=3$
conformal quantum critical systems \cite{fradkin}, as well as certain
three-dimensional (gapped) topological phases \cite{wenx}. However,
from the discussion of the present paper, it is natural to expect that
such a finite contribution will in fact depend on the details of the
state in which the EE is calculated. Certainly in the holographic
framework, this finite term should depend on
$\overset{\scriptscriptstyle{(d/2)}}{g}_{\!ij}$ and higher order terms
in the FG expansion. We have confirmed this expectation with an
explicit calculation in appendix \ref{oddu}. Hence in general, while
such a constant contribution to the EE certainly contains information
with which we may characterize the underlying field theory, it will not
be completely universal in the same sense as the coefficient of a
logarithmic contribution. In particular then, in order to properly
compare or distinguish theories with a constant contribution to EE, we
must specify that this term was calculated in the vacuum state of the
underlying theory.\footnote{In general, we would have to refine further
our characterization of these constant contributions to the
entanglement entropy. In particular, different regulators will modify
the details of the expansion the EE in terms of the cut-off and hence
ambiguities should be expected to appear in the definition of the constant
contribution.}

Of course, one of the interesting results arising from our holographic
investigations was that relevant deformations of the boundary theory
will produce new universal contributions to the EE, which are
logarithmic in the cut-off. Schematically, the general form of the
logarithmic contribution is an integral over the entangling surface
$\Sig$:
 \be
S_\mt{univ} = \sum_{i,n} \gamma_i(d,n) \int_{\Sig}\!d^{d-2}\!\sigma\,
\sqrt{H}\ [R,K]^n_i\ \mu^{d-2-2n}\,\log \mu\delta\,, \labell{EEdisk2z}
 \ee
where $n<(d-2)/2$ , $\mu$ is the mass scale appearing in the coupling of the relevant
operator, $H_{ab}$ is the induced metric on $\Sig$ and $[R,K]^n_i$
denotes various combinations of the curvatures with a combined
dimension $2n$. Both the curvature of the background geometry or the
extrinsic curvature of the entangling surface may enter these
expressions.  The universal information which would distinguish
different theories is carried in the pre-factors $\gamma_i(d,n)$.

As noted previously, for $n=0$, we have simply $[R,K]^0_1=1$ and the
integral simply yields the area of the entangling surface $\Sig$. In
this case, this contribution matches the form of the universal terms
\reef{franks} recently found for a massive free scalar \cite{frank}.
Further for $n=(d-2)/2$, the above integral involves only curvatures
(\ie the $\mu$ factor reduces to one) and our expression will match the
form found for an even-dimensional CFT \cite{rt2,solo} --- see below.
More generally, the presence of these new universal terms with $n>0$ is
easily detected with simple calculations involving symmetric
geometries, as in sections \ref{exercise} and \ref{curve}. However, the
precise form of the expressions $[R,K]^n_i$ cannot be determined in
these calculations. However, one feature that is already evident there
is that the combination $[R,K]^n_i\,\mu^{d-2-2n}$ appearing in the
integrand has dimension $d-2$, which ensures that the resulting
coefficient is scale free. With the more elaborate approach outlined in
section {pbhmatter}, the precise form of $[R,K]^n_i$ can in principle
be determined but this is a somewhat tedious exercise. Hence we have
only examined the particular case of $\alpha = (d-4)/2$ for which the
result is given in eq.~\reef{guub}. A feature of these calculations is
that rather than thinking of simply a coupling constant for the
relevant deformation, we must allow $\phi^{(0)}$ to be a field which
various over the boundary geometry. This approach also highlights the
connection of the entanglement entropy to a Graham-Witten anomaly
\cite{grwit} for the entangling surface, as noted previously for pure
CFT's in \cite{adam,EEGB}.
We should note, however, that the spacetime dimension $d$ and the
conformal dimension of the relevant operator $\Delta$ must satisfy a
particular constraint before the various terms in eq.~\reef{EEdisk2z}
can appear. These constraints simply reflect the relations
discussed at the end of section \ref{notes} where the logarithmic
contribution appears if a term in the expansion of the area
\reef{arealx} with $\tau^{n+m\frac{\alpha}{2}}=\tau^{\frac{d-2}2}$.
Hence, for a term with $n$ to appear in eq.~\reef{EEdisk2z}, we require
 \be
 \Delta=d-\frac{d-2-2n}m \qquad{\rm with\  integer}\ m\ge2\,.
 \labell{constraintzzc}
 \ee
We might note that the universal contributions from relevant
deformations typically appear in higher dimensions. The leading term
with $n=0$ appears for $d\ge3$. However, the terms mixing the
curvatures with a power of $\mu$ require larger values of $d$. For
example, we require $d\ge5$ and $d\ge7$ for the contributions with
$n=1$ and 2, respectively. Of course, as we illustrated in section
\ref{firtree2x}, that a single deformation may produce more than one of
these universal terms in higher dimensions, \ie with $d=8$ and
$\Delta=6$, eq.~\reef{constraintzzc} can be satisfied with $n=0,\,m=3$
and $n=1,\, m=2$.

Recall the integer $m$ cannot be 1 above in eq.~\reef{constraintzzc}
because the stress tensor in the bulk Einstein equations
\reef{Einstein} is at least quadratic in the scalar field. We might
compare this feature of our calculations to a similar result in
\cite{cardy6}. There, a relevant operator $\lambda\,\cal O$ is
introduced perturbatively in a two-dimensional CFT and it is noted that
this deformation only begins to have effect at order $\lambda^2$
because the one-point function $\langle\cal O\rangle$ vanishes in the
CFT vacuum. Of course, this observation extends to CFT's in any number
of dimensions and then agrees with our result that the new universal
terms appear with a factor of $\lambda^2$ or a higher power of the
coupling. However, we might also contrast the differences between the
two situations. First, in our holographic calculations, we are not
working perturbatively, \ie we are not assuming that $\lambda$ is small
in any sense. Further, one of our key observations is that the results
for the universal coefficients is independent of the state of the
boundary CFT and so does not rely on calculating in the vacuum state.
It would be interesting to see if in fact these features also extend
beyond our holographic setting to more general CFT's.

We must emphasize that eq.~\reef{EEdisk2z} is schematic. In particular,
for a given value of $n$, there may be several independent combinations
of curvatures that appear, including both background curvatures and the
extrinsic curvature of the entangling surface. Our result in
eq.~\reef{guub} explicitly illustrates the possible complications.
Further we must add that even when eq.~\reef{constraintzzc} is
satisfied, the coefficient of the universal term may still vanish,
depending on further details of the underlying theory. For example, if
the bulk scalar theory, respects a discrete symmetry $\Phi\to-\Phi$,
the coefficient will vanish unless $m$ is an even integer. It is
interesting to consider how these results would change if there were
two or more relevant deformations with different conformal dimensions.
We expect that in fact there would be no essential changes. The
asymptotic expansions would have to be extended to allow separate
factors $\rho^{\alpha_i/2}$ from each of the deformations and the
nonlinearities of the bulk theory would mix these terms. However, the
schematic structure of the universal terms would remain as given in
eq.~\reef{EEdisk2z} and the constraint on the conformal dimensions to
produce a particular term would become
 \be
\sum m_i(d-\Delta_i)=d-2-2n\,. \labell{silly}
 \ee

Of course, one contribution, which appears irrespective of the precise
conformal dimension(s) of the relevant deformation(s), is the term with
$n=(d-2)/2$. As noted above, such contributions are known to appear for
any even-dimensional CFT and the universal coefficients $\gamma$
correspond to the central charges of the CFT \cite{solo,rt2}. Even with
the relevant deformation the present case, it is precisely these terms
that appear with the central charges of the CFT that emerges in the UV
regime. As was demonstrated in \cite{EEGB}, the precise structure of
these terms and their dependence on the geometry of the boundary metric
and of the entangling surface can be derived using the PBH approach
discussed in section \ref{pbhmatter}. The latter was originally derived
in the case where the boundary theory was a pure CFT, however, all of
the same contributions still appear in the asymptotic expansion when
the relevant deformation is turned on. Hence the structure of this term
does not change in the case where the boundary theory is deformed by a
relevant deformation.

It is interesting that our holographic calculations indicate that for
even $d$, the same central charges for the CFT emerging in the UV
actually appear in the coefficients of all of the logarithmic
contributions. This can be seen from the pre-factor of $(L/\lp)^{d-1}$
which appears in all of our results. Since our bulk theory corresponds
to Einstein gravity, all of the central charges are equal and we can
not distinguish precisely which central charges appear in the various
contributions. It may be interesting to repeat our analysis in the case
where the bulk gravity theory includes higher curvature interactions
since in principle, this would allow us to distinguish the different
central charges \cite{highc} --- see below.

The appearance of central charges in these new universal contributions
hints at the close relation of these new terms with the trace anomaly.
As is well known, with even $d$, EE in a CFT can be directly calculated
using the trace anomaly, at least for geometries with sufficient
symmetry \cite{rt2,solo,EEGB,cthem2}. Typically, we consider the trace
anomaly in a curved background, where it is usually related to various
conformally invariant combinations of the curvature \cite{deser}.
However, deforming the CFT with a relevant operator will also introduce
additional terms in the trace anomaly related to the coupling to the
new operators \cite{construct,anom}. While this situation has not been
studied in detail, it is already evident that terms involving both the
curvature and the coupling of the relevant operator appear in the trace
anomaly in this situation. For example, the simplest such term, which
arises for $\Delta=(d+2)/2$, takes the form \cite{anom}
 \be
\langle\,T^i{}_i\,\rangle = {1\over 2} \phi^{(0)}\left(\Box
+\frac{d-2}{4(d-1)}\,R\right)\phi^{(0)}+\cdots\,.
 \labell{jgh}
 \ee
Given such a result, we can apply the approach, alluded to above, to
calculate the entanglement entropy using the trace anomaly
\cite{rt2,solo,EEGB,cthem2} and we have confirmed that eq.~\reef{jgh}
does indeed yield a universal contribution to the EE of precisely the
form given in eq.~\reef{EEdiskmm}. More generally, we expect that the
new universal terms in eq.~\reef{EEdisk2z} are similarly related to new
terms which the relevant operator induces in the trace anomaly. We hope
to return to these issues elsewhere. Note that the calculations which
we have sketched above apply to any general CFT with a relevant
deformation and does not refer to holography. This would demonstrate
that our results apply more broadly than to holographic field theories.

Another framework where these aspects of entanglement entropy are
easily studied is with free field theories. In particular,
ref.~\cite{frank} considered a free massive scalar in a flat background
with a flat entangling surface. They found the logarithmic terms in
eq.~\reef{franks} appear for any even $d\ge4$ and these correspond to
the $n=0$ terms in the general expression \reef{EEdisk2z}. It is
amusing to compare this result to our holographic results which would
correspond to a strongly coupled field theory. The natural relevant
operator to consider would be one with $\Delta=d-2$, as for a scalar
mass term. In this case, the holographic contribution appears again for
any even $d$ but for $d\ge 6$. One can easily extend the free scalar
calculation to examples where the geometry is curved \cite{unpub} and
logarithmic terms mixing curvatures and powers of the mass also appear
in higher dimensions, similar to the results of our holographic study.
Another simple extension of the free field calculations in \cite{frank}
is to consider massive fermions \cite{unpub}. In this case, it appears
that various logarithmic contributions as in eq.~\reef{EEdisk2z} again
arise in even dimensions. On the holographic side, it would be natural
to compare to a relevant operator with $\Delta=d-1$, as for a fermionic
mass term. In this case, the holographic calculations yield a
logarithmic contribution for any odd or even dimension with $d\ge4$.
Hence this discussion again indicates that the new universal terms
\reef{EEdisk2z}, which we have uncovered here with holographic
calculations, have a broader applicability and also arise in
calculations of EE for more conventional field theories.

To close, let us observe that our analysis always assumed that the bulk
gravitational theory was simply Einstein gravity and that the
holographic EE was given by the standard Ryu-Takayanagi proposal
\reef{define}. In various contexts, it would be interesting to consider
the addition of higher curvature interactions to the bulk theory. The
modification that such interactions would make in the holographic EE is
not completely resolved. It is expected that eq.~\reef{define} would be
replaced by the extremization of some geometric functional which
produces the correct black hole entropy when evaluated on an event
horizon. Recent progress was made in this direction for Lovelock
theories of gravity \cite{EEGB,friends}. For example, with Gauss-Bonnet
gravity in the bulk, eq.~\reef{define} is replaced by the following:
 \beq
S(V) = \frac{2 \pi}{\lp^{d-1}}\ \mathrel{\mathop {\rm
ext}_{\scriptscriptstyle{\m\sim V}} {}\!\!}  \int_\m
d^{d-1}\!y\,\sqrt{h}\,\left[ 1+ \lambda\,L^{2}\,\R \right]\,,
 \labell{Waldformula3x}
 \eeq
where $\R$ denotes the Ricci scalar for the intrinsic geometry on $\m$
and $\lambda$ is the (dimensionless) coupling which controls the
strength of the curvature-squared interaction. While the appropriate
entropy functional is not known for general higher curvature theories,
we expect that it will have the form
 \be
S_{EE}=\frac{2\pi}{\lp^{d-1}}\ \mathrel{\mathop {\rm
ext}_{\scriptscriptstyle{\m\sim V}} {}\!\!} \int_\m
d^{d-1}\!y\,\sqrt{h}\,\left[ 1+ f(R,K^i,\Phi)\right]\,,
 \labell{wild}
 \ee
where $f$ is some local scalar constructed from the bulk curvature $R$,
the extrinsic curvature of the surface $K^i$, the bulk scalar $\Phi$
(when a relevant deformation is introduced) and derivatives of these
building blocks. Note that eq.~\reef{Waldformula3x} can be re-expressed
in this form using the Gauss-Codazzi equations \cite{EEGB}. Now the key
observation is that in the FG gauge, any such scalar will admit an
expansion in $\tau$ beginning at order $\tau^0$. Hence the expansion of
the full integrand begins with $\tau^{-\frac{d}{2}}$, just as in
section \ref{notes}. Further, the PBH transformations  will continue to
fix the asymptotic form of the asymptotic metric (and scalar), as well
as the embedding functions, as discussed in section \ref{pbhmatter}.
The only change is that the constants appearing at various orders may
take on new values as the equations of motion will have changed. In any
event, as in the main text, any logarithmic contribution will only
depend on the fixed boundary data and so we expect that the
corresponding coefficient remains state independent when the bulk
gravity theory is extended to include higher curvature interactions.
Hence our previous result for the universal logarithmic contribution to
the holographic EE is not changed in such a generalized holographic
framework. Further we do not expect that the basic form of the
universal terms will change in this scenario. However, it may be useful
to examine these expressions in, \eg Lovelock gravity, as it may allow
one to identify the specific central charge associated with the pre-factor
$(L/\lp)^{d-1}$, as discussed above.

\acknowledgments

We would like to thank Alex Buchel, Joao Penedones, Mark van Raamsdonk, 
Simon Ross, Brian Swingle and especially Horacio Casini for useful
conversations and correspondence. Research at Perimeter Institute is
supported by the Government of Canada through Industry Canada and by
the Province of Ontario through the Ministry of Research \& Innovation.
RCM also acknowledges support from an NSERC Discovery grant and funding
from the Canadian Institute for Advanced Research. RCM would also like
to thank the Aspen Center for Physics in the final stages of preparing
this paper.

\appendix

\section{Universality and Odd $d$} \label{oddu}

In many cases, no logarithmic contribution appears in the EE, \eg with
a CFT in an odd number of spacetime dimensions. However, the constant
term appearing in usual expansion in powers of the short-distance
cut-off may still be regarded as a universal contribution to the EE
\cite{rt2}. The universality of this constant contribution is
well-established for a variety of $d=3$ conformal quantum critical
systems \cite{fradkin}, as well as certain three-dimensional (gapped)
topological phases \cite{wenx}. However, from the discussion of the
present paper, it is natural to expect that such a finite contribution
will in fact depend on the details of the state in which the EE is
calculated. Certainly in the holographic framework, this finite term
should depend on $\overset{\scriptscriptstyle{(d/2)}}{g}_{ij}$ and
higher order terms in the FG expansion. Hence in general, if we are to
interpret this constant contribution to the EE as characteristic of the
underlying field theory, we must also specify that the calculations
were performed in the vacuum state of the theory.

In the following, we verify that the finite term in the EE does in fact
depend on the state of the underlying theory with a simple holographic
calculation. We consider a boundary CFT at finite temperature $T$ and
calculate the holographic EE across a spherical entangling surface of
radius $R$. Working at low temperature, \ie $RT\ll1$, we identify a
temperature dependent contribution to the finite term in the EE. Note
that this contribution is not simply the entropy density of the thermal
bath multiplied by the volume of the ball bounded by the sphere --- we
comment on this point at the end of the appendix.

For simplicity, we consider the CFT in a flat $d$-dimensional spacetime
and so at finite temperature, the holographic dual is a
$(d+1)$-dimensional planar AdS black hole. The metric for this bulk
solution can be written as
 \be
ds^2={L^2 \over z^2} \left( -f(z)\,dt^2+dr^2+r^2d\Omega^2_{d-2}
\right)+{L^2\over f(z)}{dz^2 \over z^2}~,
 \labell{planarBH}
 \ee
where we have introduced polar coordinates in the boundary directions
and $f(z)$ is given by
 \be
f(z)=1-\Big({z \over z_+}\Big)^d~.
 \ee
Note that in this solution, the horizon appears at $z=z_+$. Further, in
the limit $z_+ \to \infty$, we recover the AdS vacuum metric in
Poincare coordinates. The Hawking temperature of this black hole
solution is given by
 \be
 T={d \over 4\pi z_+}~.
 \labell{BHtemp}
 \ee

To evaluate the holographic EE \reef{define} for a spherical entangling
surface, we must determine the extremal bulk surface described by a
profile $r=r(z)$ with the boundary condition $r(z=0) =R$, as in section
\ref{sec:disk} except that the bulk space is now given by
eq.~\reef{planarBH}. The induced metric on such a surface is given by
 \be
h_{\al \bt}\,dx^{\al}dx^{\bt}=\frac{L^2}{z^{2}}\left[ ( {r'}^2+
1/f)\,dz^2+r^2d\Omega^2_{d-2}\right]
 \quad,
 \labell{eqn:ind-metric-diska}
 \ee
where the `prime' denotes a derivative with respect to $z$. As a result
the EE is given by
 \be
S=2\pi\frac{L^{d-1}}{\lp^{d-1}}\,\Omega_{d-2}\int dz \, {r^{d-2} \over
z^{d-1}}\sqrt{r'^2+1/f} ~.
 \labell{EEdiskT}
 \ee

In the case of pure AdS ($z_+ = \infty$) the shape of the extremal
surface can be obtained in the closed form given in eq.~\reef{soluble}
\cite{rt1,rt2}. However, for general $z_+$ we did not succeed to find a
closed analytic expression and thus we proceed perturbatively in $R/z_+
<< 1$. In light of eq.~\reef{BHtemp}, this regime can be interpreted as
a low temperature limit $TR <<1$ (for a fixed number of dimensions
$d$). In this regime, the extremal surface will be close to that in
eq.~\reef{soluble} and so we have $z\lesssim R$. Therefore we expand
the integrand of eq.~\reef{EEdiskT} in powers of
$\epsilon(z)=(z/z_+)^d$
 \be
S=2\pi\frac{L^{d-1}}{\lp^{d-1}}\,\Omega_{d-2}\int_{\delta}^{z_{max}} dz
\, {r^{d-2} \over z^{d-1}}\sqrt{ r'^2+1} \Big(1+{\epsilon(z) \over2(
r'^2+1)}+\mathcal{O}(\epsilon^2)\Big) \,.
 \labell{EEdiskT1}
 \ee

Recall that with $z_+=\infty$, $\epsilon(z)=0$ and, according to
eq.~\reef{soluble} \reef{soluble}, the profile of the extremal surface
is given by $r_0(z)=\sqrt{R^2-z^2}$ and $z_{max}=R$. However, with
$\epsilon(z)\neq 0$, both $r(z)$ and $z_{max}$ acquire corrections
 \be
r(z)=r_0(z)+\delta r(z)~, \quad z_{max}=R+\delta z_{max}~,
 \labell{changes}
 \ee
where these corrections are at least of order $\epsilon$. To solve for
$\delta r(z)$, we may substitute eq.~\reef{changes} into the action
\reef{EEdiskT1} and consider extremizing with respect to $\delta r(z)$.
However, in doing so, we find that to leading order there are two
contributions, one of order $\delta r^2$ and the other of order
$\epsilon\,\delta r$. Hence upon substituting the solution back into
eq.~\reef{EEdiskT1}, we would find that to leading order $\delta r$
only makes contributions of $O(\epsilon^2)$ and so we can ignore this
change in the profile. Similarly, the contribution to
eq.~\reef{EEdiskT1} from the change $\delta z_{max}$ involves
evaluating the integrand (with $r=r_0$) at $z_{max}=R$ but this
vanishes to leading order since $r_0(R)=0$. Therefore if we work only
to linear order in $\epsilon(z)$, we need only evaluate the second term
in eq.~\reef{EEdiskT1} with the profile $r_0(z)$:
 \be
\delta S=2\pi\frac{L^{d-1}}{\lp^{d-1}}\,\Omega_{d-2}\int_{\delta}^{R}
dz \, {r_0^{d-2} \over z^{d-1}}{\epsilon(z)
\over2\sqrt{r_0'^{\,2}+1}}+\mathcal{O}(\epsilon^2)=
2\pi\frac{L^{d-1}}{\lp^{d-1}}\,{\Omega_{d-2} \over 2(d+1)}\Big( {R\over
z_+} \Big)^d+\mathcal{O}(\delta^d,\epsilon^2)~.
 \labell{EEdiskT2}
 \ee
Hence combining this result with eq.~\reef{BHtemp}, we find that
$\delta S\sim (RT)^d$ and so we have found a finite contribution to the
EE which depends on the temperature (state) of the boundary CFT. Let us
also note that for two-dimensional CFT's, one can get a closed
expression for the EE at finite temperature \cite{cardy0} and expanding
the latter in the limit of low temperature reproduces precisely our
correction \reef{EEdiskT2}.

Let us consider extending the above calculation to a more general
entangling surface to provide a general estimate for the contribution
$\delta S$ calculated above for a spherical surface. To make progress
here, it is simplest to adopt the general framework and notation
introduced in section \ref{notes}. In particular, we begin by adopting
the usual radial coordinate of the FG expansion \reef{expandfg}
 \bea
z&=&L \rho^{1/2} \left( 1+ \frac14 \left(\frac{L}{z_+}\right)^d
\rho^{d/2} \right)^{-2/d}\nonumber\\
& \simeq& L\,\rho^{1/2}\left(1-\frac{1}{2d}
\left(\frac{L}{z_+}\right)^d \rho^{d/2}+\cdots\right)\,.
 \labell{FGradius}
 \eea
With this choice, the asymptotic expansion of the planar black hole
metric \reef{planarBH} becomes
 \be
ds^2\simeq{L^2\over 4\rho^2}\,{d\rho^2}+{1 \over \rho} \left[ -\left(
1-\frac{d-1}{d} \left(\frac{L}{z_+}\right)^d
\rho^{d/2}\right)dt^2+\left( 1+\frac{1}{d} \left(\frac{L}{z_+}\right)^d
\rho^{d/2}\right)\sum (dx^i)^2 \right]~.
 \labell{planarBH2}
 \ee
Hence as expected, we see that the leading effect of the temperature
appears in $\overset{\scriptscriptstyle{(d/2)}}{g}_{ij}$ in the FG
expansion \reef{expand}. Now we choose some entangling surface $\Sig$
in the flat boundary metric which is described by
$\overset{\scriptscriptstyle{(0)}}{X}{}^i(y^a)$ and there will be some
bulk surface described by the profile $X^i(y^a,\tau)$ satisfying the
boundary condition $X^i(y^a,\tau=0)=
\overset{\scriptscriptstyle{(0)}}{X}{}^i(y^a)$. We make the same gauge
choice as in eq.~\reef{eqn:gauge-brane} and then the induced metric
\reef{induceh} becomes
 \bea
h_{\tau\tau}&=&\frac{L^2}{4\tau^2}\left(1+\frac{4\tau}{L^2}
\partial_\tau X^i \partial_\tau X^j\,g_{ij}\right)\equiv \frac{L^2}{4\tau^2}
\tilde{h}_{\tau\tau}\,,\nonumber\\
h_{ab}&=&\frac{1}{\tau}\,\partial_a X^i \partial_b X^j \,g_{ij}\equiv
\frac{1}{\tau} \tilde{h}_{ab}\,.\label{induceh33}
 \eea
As only the spatial coordinates are relevant here, we may write
 \be
g_{ij}\simeq \delta_{ij}\left(1+\tilde{\epsilon}(\rho)\right) \quad{\rm
where}\ \ \tilde{\epsilon}(\rho)=\frac{1}{d}\left( \frac{L}{z_+}
\right)^d \rho^{d/2}\,.
 \labell{deltag}
 \ee

Now following the analysis above for the spherical entangling surface,
we will expand the holographic EE in powers of $\tilde{\epsilon}(\rho)$
and only keep the contribution that is linear in this term. As above,
the deformation of the background metric will produce perturbations of
the profile and the maximum value of $\rho$, both of which begin at
linear order in $\tilde{\epsilon}(\rho)$:
 \be
X^i(y^a,\tau)=X_0^i(y^a,\tau)+\delta X^i(y^a,\tau)~, \quad
\rho_{max}=\rho_{0,max}+\delta \rho_{max}~.
 \labell{changes2}
 \ee
Again, to solve for $\delta X^i(y^a,\tau)$, we would extremize the area
functional with respect to these functions. However, also as above, we
find that to leading order there are two contributions, one of order
${\delta X^i}^2$ and the other of order $\epsilon\,\delta X^i$. Hence
upon substituting the solution back into area, we would find that to
leading order $\delta X^i$ only makes contributions of $O(\epsilon^2)$
and so we can ignore these changes in the profile. Similarly, the
contribution from the change $\delta \rho_{max}$ involves evaluating
the integrand with $X^i_0$ at $\rho_{0,max}$ but this vanishes (to
leading order) since by definition $\rho_{0,max}$ is the point where
the bulk surface (smoothly) closes off. Hence at this point, we have
$\sqrt{h_0}|_{\rho=\rho_{0,max}}=0$. Therefore if we work only to
linear order in $\epsilon(z)$, we need only evaluate the variation to
the area coming from the change of the background metric, given in
eq.~\reef{deltag}, with the profile $X^i_0$:
 \bea
\delta S&=&\frac{2\pi}{\lp^{d-1}}\oint_\Sig
d^{d-2}y\int_{\delta}^{\rho_{max}} d\tau \,
\frac{L}{2\,\tau^{d/2}}\sqrt{\tilde{h}_0}\, \left[\frac12
\tilde{h}_0^{\al\bt}\, \delta_{\tilde{\epsilon}} \tilde{h}_{\al\bt}
\right]
 \nonumber \\
&=&\frac{2\pi}{\lp^{d-1}}\oint_\Sig d^{d-2}y\int_{\delta}^{\rho_{max}}
d\tau \, \frac{L}{4\,d}\sqrt{\tilde{h}_0} \left[ d-1 -
\frac{1}{(\tilde{h}_0)_{\tau\tau}} \right]
\,\left(\frac{L}{z_+}\right)^d\,.
 \labell{deltaEE}
 \eea

An essential feature of this result is that it is finite, \ie
the leading factor of $\tau^{-d/2}$ has been canceled by the
$\tau$-dependence of $\tilde{\epsilon}$. In the vicinity of $\tau=0$,
the integrand reduces to essential the $\sqrt{{\rm
det}\overset{\scriptscriptstyle{(0)}}{h}_{ab}}$, \ie the area measure
on $\Sig$ in the boundary metric. Hence we would argue that the $y$
integration essentially contributes a factor of $\cA_{d-2}$, the area
of the entangling surface. Certainly such a factor appears for
entangling surfaces with sufficient symmetry, such as the spherical
surface in the previous calculation. Similarly, the contribution of the
$\tau$ integral can be estimated to be roughly $\rho_{max}\simeq
\ell^2/L^2$, where $\ell$ is some characteristic scale of the geometry
of $\Sig$ which controls how far the extremal surface extends into the
bulk. Again, for surfaces with sufficient symmetry, we can readily
identify $\ell$. For example, in the above calculations, $\ell$ is the
radius $R$ of the spherical entangling surface or in section
\ref{sec:belt}, $\ell$ would be the width of the slab with flat
parallel boundaries. Therefore up to overall numerical factors, our
estimate of this contribution to the holographic EE becomes
 \be
\delta S\simeq \frac{L^{d-1}}{\lp^{d-1}}\,\cA_{d-2}\,\ell^2\,T^d\ .
 \labell{deltaEE3}
 \ee

Thus, our holographic calculations explicitly demonstrate that the
constant contribution to the EE depends on the state in which the
latter is calculated. Hence, while such a contribution certainly
contains information that characterizes the underlying field theory, we
must be careful in comparing various results to specify the state (\eg
the vacuum) for which the calculations were performed.

Let us consider our holographic result \reef{deltaEE3} further. Given
that the boundary CFT is at finite temperature $T$, the thermal bath
will produce a uniform entropy density $s \sim
(L^{d-1}/\lp^{d-1})\,T^{d-1}$ and so for a general region with volume
$\cV_{d-1}$, the corresponding thermal entropy would be $\delta
S_{therm} = (L^{d-1}/\lp^{d-1})\,\cV_{d-1}\,T^{d-1}$. Hence comparing
this result to eq.~\reef{deltaEE3}, we see that the finite temperature
dependent contribution to the holographic EE which we have identified
does not correspond to this thermal entropy. It may seem that we have
found a discrepancy since we should expect that $\delta S_{therm}$
should appear as a finite contribution in the EE \cite{rt1,rt2} and in
fact, it seems that this contribution would dominate in the low
temperature limit (given that $\delta S_{therm}$ is proportional to a
smaller power of $T$). However, the latter limit provides the
resolution of this apparent discrepancy. Since we are working in the
limit $\ell T\ll1$, the typical wavelength of the thermal excitations
is much larger than the size of the entangling geometry and so it
should not be a surprise that $\delta S_{therm}$ has not appeared in
our calculations. Instead this contribution would be expected to appear
in the opposite limit $\ell T\gg1$. In the latter case, the bulk
surface would extend down to event horizon at $z=z_+$ and $\delta
S_{therm}$ would naturally be produced by the portion of the extremal
surface stretched along the horizon.

\end{document}